\documentclass[5p,times,omitsubmissiontext]{cas-dc}
\usepackage[numbers,sort&compress]{natbib}
\usepackage{flushend}
\usepackage{setspace}
\usepackage{enumitem}
\usepackage{csquotes}
\usepackage{stfloats}
\usepackage{graphicx}  
\usepackage{tabto}
\usepackage{supertabular}  
\usepackage{times}  
\usepackage{caption}
\usepackage{float}
\usepackage{pdflscape}   
\usepackage{lscape}      
\usepackage{textcomp} 
\usepackage{booktabs}   
\usepackage{multirow}   
\usepackage{graphicx}   
\usepackage{geometry}
\usepackage{fancyhdr}
\DeclareUnicodeCharacter{2212}{-} 
\def\tsc#1{\csdef{#1}{\textsc{\lowercase{#1}}\xspace}} 
\tsc{WGM}
\tsc{QE}
\tsc{EP}
\tsc{PMS}
\tsc{BEC} 
\tsc{DE}

\makeatletter
\def\ps@first{%
  \let\@oddhead\@empty
  \let\@evenhead\@empty
  \def\@oddfoot{} 
  \let\@evenfoot\@oddfoot
}
\makeatother

\begin{document}
\singlespacing
\let\WriteBookmarks\relax
\def\floatpagepagefraction{1}
\def\textpagefraction{.001}

\newgeometry{
    top=2cm,
    bottom=2cm,
    left=1.5cm,
    right=1.75cm,
    footskip=0pt
}


\shortauthors{Rathinaraja Jeyaraj et~al.}


\title[mode=title]{%
    \texorpdfstring{\includegraphics[width=0.65cm]{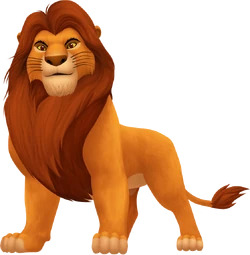}\hspace{0.25em}}{}%
    \textbf{MUFASA: An Information Utility-Aware Preprocessing Framework for Reliable Model Reasoning in Computational Pathology}%
}

\author[1]{Rathinaraja Jeyaraj}

\author[1]{Barathi Subramanian}  

\author[1,2]{Songmi Noh}\fnmark[$\dagger$]

\author[1]{Mitchell N. Peterson}  

\author[1]{Terry Guo}  

\author[1]{George A. Fisher}  

\author[3,4]{Nigam Shah}  

\author[4,5]{Curtis P. Langlotz}  

\author[1]{Thomas J. Montine}  

\author[1,4]{Jeanne Shen}\cormark[1]
\ead{jeannes@stanford.edu}

\cortext[1]{Corresponding author, $\dagger$Present address}

\affiliation[1]{organization={Department of Pathology, Stanford University School of Medicine},
    city={Stanford},
    postcode={94305}, 
    state={CA},
    country={USA}}

\affiliation[2]{organization={Department of Pathology, CHA Gangnam Medical Center, CHA University School of Medicine},
    city={Seoul},
    postcode={13488}, 
    country={South Korea}}

\affiliation[3]{organization={Department of Medicine, Stanford University School of Medicine},
    city={Stanford},
    postcode={94305}, 
    state={CA},
    country={USA}}

\affiliation[4]{organization={Center for Artificial Intelligence in Medicine \& Imaging},
    city={Stanford},
    postcode={94305}, 
    state={CA},
    country={USA}}

\affiliation[5]{organization={Department of Radiology, Stanford University School of Medicine},
    city={Stanford},
    postcode={94305}, 
    state={CA},
    country={USA}}

\begin{abstract}
Reliable computational pathology depends on preprocessing methods that identify informative tissue regions while excluding artifacts and low-utility regions from whole-slide images (WSI). However, existing preprocessing pipelines often retain such regions or discard diagnostically relevant tissue, thereby limiting downstream model performance, reliability, and robustness across heterogeneous cohorts. Here, we systematically evaluate how these regions affect downstream AI model performance across multiple clinically relevant applications and introduce MUFASA, a generalizable, information utility-aware preprocessing framework for H\&E-stained WSI that excludes artifacts and low-utility regions while preserving biologically meaningful tissue. MUFASA integrates slide-level artifact masking, stain-aware tile filtering, reconstruction-based utility stratification of tiles, and targeted recovery of tissue tiles that are over-filtered by earlier phases. Across tumor diagnosis, tumor subtyping, biomarker status prediction, and survival prognostication tasks in diverse cancer cohorts, MUFASA consistently improves downstream model performance relative to widely used preprocessing baselines. These gains are accompanied by reduced artifact-associated attribution in model heatmaps, indicating improved alignment between retained tissue and model attention. Our findings establish WSI preprocessing as a critical determinant of downstream model performance and validity, revealing that even accurate predictions can conceal important failure modes stemming from anatomically implausible reasoning driven by retained artifact-containing and low information-utility tiles. 
\end{abstract}



\maketitle
\section{Introduction}
\begin{figure*}[t!] 
	\centering 
	\includegraphics[width=0.99\textwidth]{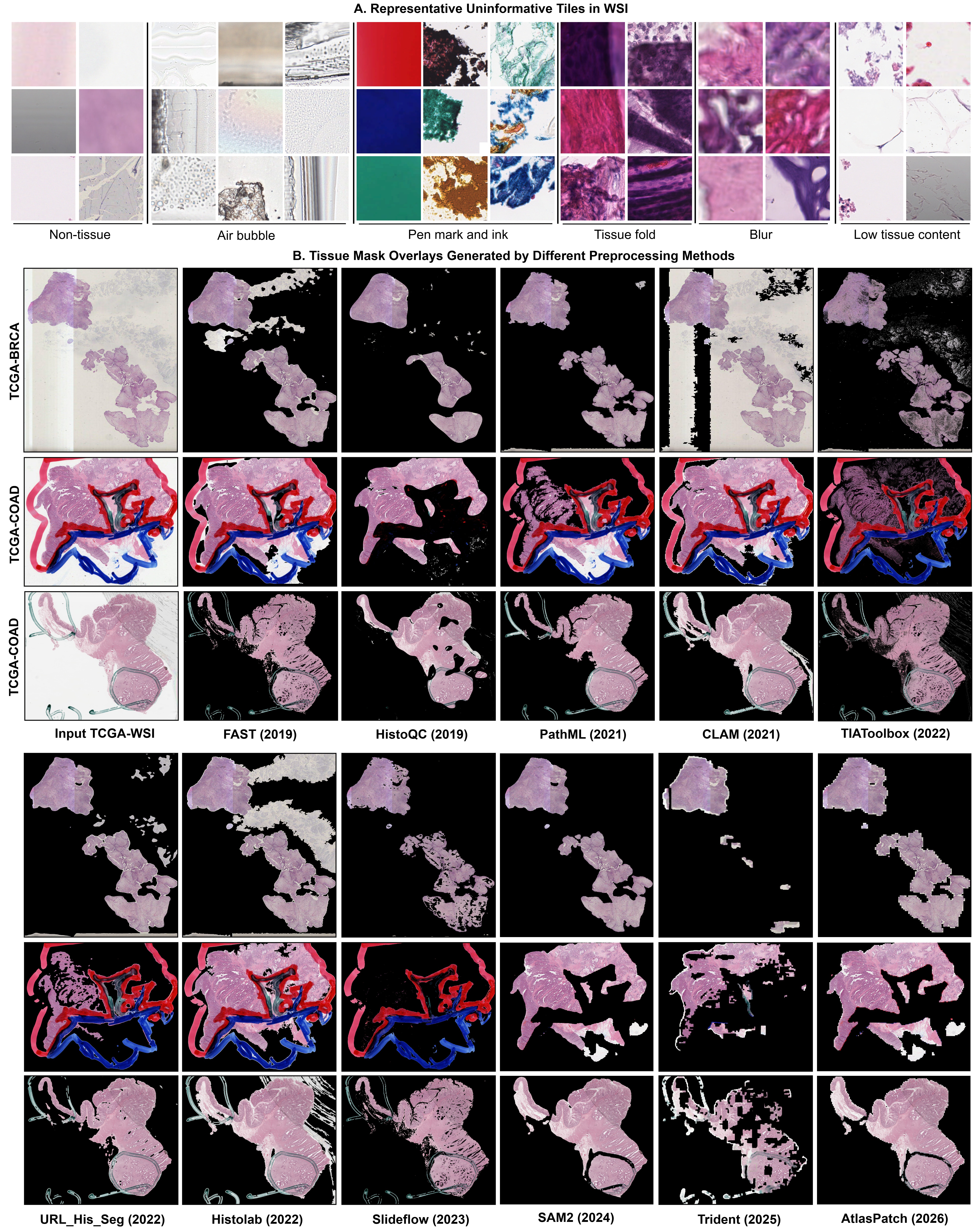}
	\\
	\caption{\footnotesize \textrm{\textbf{Uninformative tiles and preprocessing outputs from H\&E-stained whole-slide images (WSI).} 
    \textbf{A}. Representative tile-level examples of non-tissue background, common artifacts, and low tissue-content regions. 
    \textbf{B}. Qualitative comparison of tissue-mask overlays produced by widely used WSI preprocessing methods (year of publication in parentheses) on three representative WSI from the TCGA-COAD and TCGA-BRCA cohorts illustrating differences in artifact retention and tissue loss. }}
	\label{fig:1}       
\end{figure*}

Computational pathology (CPath) \cite{1,2} has emerged as a transformative frontier in precision medicine, enabling the scalable analysis of hematoxylin and eosin (H\&E)-stained whole-slide images (WSI) for diagnostic classification \cite{3}, segmentation \cite{4,5}, biomarker inference \cite{6}, and prognostic modeling \cite{7}, among a multitude of other applications. In most CPath pipelines, WSI are divided into fixed-size tiles, deep encoders extract tile-level features, and multiple instance learning (MIL) models aggregate these features into slide- or patient-level predictions. Although MIL-based models have shown promising performance across diverse clinical endpoints, their reliability depends critically on the quality and relevance of tissue regions extracted from each slide \cite{8}. In routine practice, WSI contain a substantial proportion of uninformative regions \cite{8,9,10,11}, including non-tissue background and low tissue-content regions and common artifacts such as pen marks, obscuring debris, air bubbles, tissue folds, blur, scanner-induced distortion/artifacts, and stain artifacts (over- and under-staining) (Fig.~\ref{fig:1}A), which are introduced during tissue processing, slide preparation, digitization, and diagnostic review. When such regions are retained during tiling (e.g., subdivision of whole-slide images into smaller image tiles), they can distort learned representations, encouraging reliance on spurious visual cues and shifting model attribution away from biologically meaningful tissue morphology \cite{9,12}, even when slide-level predictions appear to be correct. This is particularly problematic for prognostic tasks, where outcome-relevant signals may be sparse and susceptible to artifacts. In addition, tiles with limited discriminative morphologic content, such as partially tissue-filled regions and background-dominant tissue regions \cite{13} (adipose-like tissue), provide low information content and can further dilute learning and divert model attention away from clinically relevant features. Robust tissue identification is therefore a fundamental prerequisite for reliable CPath, as it ensures that downstream models are trained on biologically meaningful signals rather than artifact-driven correlations, thereby improving predictive performance and model interpretability.

However, separating informative tissue from background and artifacts in H\&E-stained WSI remains challenging across heterogeneous clinical cohorts. Classical heuristic approaches, such as global Otsu thresholding \cite{14} and entropy-based selection \cite{15}, are computationally efficient but rely on fixed rules that are sensitive to stain variation, scanner noise, and background complexity, frequently misclassifying high-contrast artifacts as tissue. Widely used preprocessing pipelines such as CLAM \cite{16}, Trident \cite{17}, and Histolab \cite{18} often combine heuristic tissue masking with additional filtering, which can improve coarse localization but leave fine-grained artifacts embedded within tissue regions. Conversely, more stringent filtering \cite{17} may over-segment tissue and discard diagnostically relevant regions, sometimes producing near-empty tile sets. Deep learning-based tissue segmentation methods \cite{19,10,15,20,21}, including URL\_His\_TissueSeg \cite{22}, SAM2 \cite{23}, AtlasPatch \cite{24}, WSI-SmartTiling \cite{25}, and U-Net variants \cite{26,27}, can generate cleaner thumbnail-level tissue masks, but large-scale deployment is often constrained by annotation burden, computational cost, and inconsistent robustness across cohorts. In addition, these often require post-processing to correct intra-tissue holes \cite{19,28} (background islands present within tissue) and remove missed artifact regions \cite{17}. For instance, a GAN-based inpainting approach \cite{29} has been employed to digitally reconstruct underlying tissue morphology within pen-marked regions. Moreover, sample-selection methods \cite{13,30,31,32} applied after preprocessing are primarily intended to reduce computational cost by prioritizing diagnostically relevant or informative regions using similarity- or statistics-based criteria. However, because these strategies are often unsupervised or task-specific, residual artifacts may still propagate into downstream model inputs, and the selected samples may exhibit limited generalizability across tasks and cohorts. 

Tile-level classification approaches \cite{33,19,12,10,34} tend to reduce preprocessing to a binary tissue-versus-artifact decision, despite the reality that tile "utility" is inherently continuous and task-dependent. However, visually similar artifacts and tissue structures can be difficult to distinguish under domain shift \cite{33}, limiting generalizability. Quality control and preprocessing toolkits, such as SliDL \cite{35}, GrandQC \cite{36,37}, FAST \cite{38}, HistoQC \cite{39}, PathML \cite{40}, TIAToolbox \cite{41}, and Slideflow \cite{42}, provide structured workflows for identifying uninformative regions, but they typically prioritize either rule-based filtering or discrete artifact detection rather than utility-aware selection of tiles optimized for downstream learning objectives. As a result, current preprocessing methods face a persistent trade-off: permissive pipelines preserve tissue at the cost of artifact leakage, whereas stringent pipelines suppress artifacts but risk removing subtle yet clinically meaningful compartments, including lightly stained, mucin-rich, or adipose-rich tissue (Fig.~\ref{fig:1}B). Collectively, these limitations highlight that optimization of WSI preprocessing remains a critical yet under-addressed challenge in CPath, despite its universal impact on the performance and reliability of downstream AI models for biomedical applications. Because model behavior is inherently constrained by input data quality, preprocessing directly shapes both predictive accuracy and the biological validity of model interpretation.     

\begin{figure*}[t!] 
	\centering 
	\includegraphics[width=0.95\textwidth]{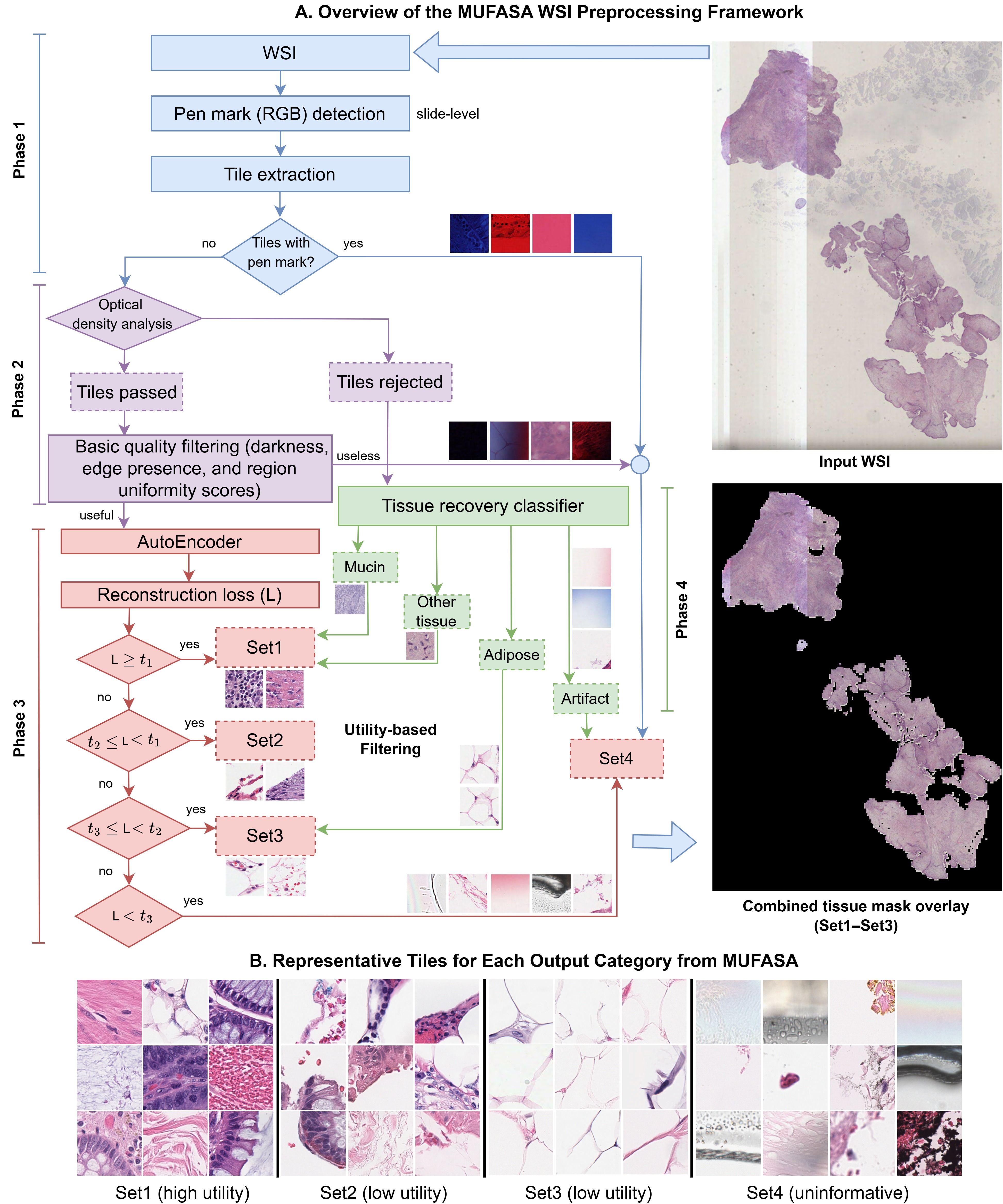}
	\\
	\caption{\footnotesize \textrm{\textbf{MUFASA preprocessing framework and representative output tile categories. }
    \textbf{A}. Schematic overview of the four-phase MUFASA pipeline. Phase 1: detection and masking of background and pen marks directly on the WSI prior to tile extraction. Phase 2: application of optical-density-based filtering to identify faintly stained tiles, followed by lightweight quality checks to remove tiles dominated by background, feature-poor regions lacking discernible tissue morphology (for example, blurry regions), and dark regions commonly associated with tissue folds, ink residue, or stain-related artifacts. Phase 3: unsupervised autoencoder-based utility stratification in which retained tiles from Phase 2 are grouped by reconstruction loss into Set1 (highly informative, tissue-dominant tiles with clear histological structure), Set2 (tiles containing partial tissue content), Set3 (low-density or subtle tissue compartments, typically from adipose-containing regions), and Set4 (uninformative tiles dominated by non-tissue or artifact-related content). Phase 4: targeted recovery of visually subtle but biologically relevant tissue compartments, such as adipose- and mucin-rich tiles removed by Phase 2. Retained tiles are remapped to their original WSI to generate interpretable overlays and aggregate tissue masks for downstream analysis. A representative input WSI from TCGA-BRCA and its corresponding preprocessed tissue-mask overlay (Set1-Set3) are shown. 
    \textbf{B}. Representative tiles from each MUFASA output category.  }}
	\label{fig:2}       
\end{figure*} 

To address these limitations, we introduce MUFASA (Multi-phase information Utility-aware Filtering and Artifact Suppression Architecture), an AI-driven preprocessing framework that robustly identifies informative tissue and generates reliable WSI-level tissue masks from H\&E-stained WSI. MUFASA is designed to suppress recurrent artifacts (Fig.~\ref{fig:1}A), without relying on a single hand-tuned heuristic or a rigid segmentation rule. Instead, MUFASA integrates four complementary phases (Fig.~\ref{fig:2}A) that progressively remove uninformative regions, such as non-tissue background and artifacts, as well as low-utility regions, including partially tissue-filled and background dominant regions, while preserving biologically meaningful tissue regions. In Phase 1, pen marks are detected and masked at the whole-slide level before tiling. Tiles overlapping pen-marked regions are classified as uninformative and assigned to Set4, whereas tiles from the remaining slide area are passed to the next processing phase. Phase 2 applies optical density-based filtering and lightweight quality checks to discard additional uninformative tiles (for example, those comprising non-tissue background, blur, and staining artifacts). Phase 3 introduces an unsupervised autoencoder-based utility model that stratifies retained tiles into four groups according to reconstruction loss (L) using thresholds ($t_1$, $t_2$, $t_3$), enabling information utility-aware filtering beyond binary “tissue versus artifact” decisions. 

As shown in Fig.~\ref{fig:2}B, tiles are ultimately grouped according to information utility into Set1, representing high-utility, tissue-rich regions with discriminative morphology; Set2, representing low-utility partially tissue-filled regions; Set3, representing low-utility background-dominant regions with limited discriminative morphology, including adipose-rich tissue; and Set4, representing no-utility or uninformative regions, including non-tissue background, artifact-containing tiles, and low tissue-content regions. Phase 4 optionally recovers biologically relevant but visually low-contrast tissue compartments, such as adipose- and mucin-rich regions, that might have been removed inadvertently during Phase 2 due to faint staining or a background-like appearance. The retained tiles are subsequently remapped to their original spatial coordinates to generate WSI-level tissue masks, along with set-specific overlays, for downstream modeling and visualization. This decomposition enables transparent inspection of tissue quality and preserves spatial context while maintaining flexibility for WSI-level tasks and broader applications, such as dataset curation and tissue-composition analyses. For the downstream WSI-level analyses in this study, only tiles from Set1 were retained, as they provide the most substantial task-relevant, high-utility signal from biologically meaningful tissue. In contrast, tiles in Set2 and Set3 were disregarded as relatively low-utility, since partially tissue-filled regions and adipose-containing regions typically offer limited discriminative morphology and weaker task-relevant signal. Tiles in the uninformative set were excluded from downstream analysis because they were unlikely to contain meaningful biological signal and might instead introduce noise into model training and evaluation. A detailed description of MUFASA is provided in the Methods section.

We evaluated MUFASA's robustness and generalizability across four clinically relevant WSI-level tasks: tumor diagnosis and subtyping, biomarker status prediction, and survival prognostication, using The Cancer Genome Atlas (TCGA) \cite{43}, CAMELYON16 \cite{44}, and an in-house Stanford colorectal cancer (CRC) cohort, spanning diverse organs and acquisition conditions. Beyond introducing MUFASA as an information utility-aware preprocessing framework, our study makes three additional important contributions. First, it provides a systematic quantitative evaluation of widely used preprocessing pipelines, demonstrating that retention of low-utility and uninformative tiles and removal of high-utility tissue regions have measurable consequences for downstream classification, biomarker status prediction, survival modeling, and model attribution across multiple MIL architectures. Second, it reveals that existing preprocessing approaches can cause models to rely on biologically irrelevant regions, and that even correct slide-level predictions may arise from spurious associations which reduce model interpretability and generalizability. In particular, our attention-based analyses reveal that several state-of-the-art CPath pipelines assign substantial importance to low-utility or uninformative regions, indicating that apparent predictive success does not necessarily reflect biologically credible reasoning. Third, we perform expert analysis of tissue-versus-artifact trade-offs to confirm whether the performance gains observed with MUFASA were accompanied by improved tile quality and more reliable preservation of diagnostically relevant tissue. By jointly quantifying these failure modes and providing a solution which optimizes the balance between uninformative region exclusion and tissue retention, our study establishes preprocessing as a critical determinant of downstream CPath model performance, rather than just a routine upstream step. To support reproducible benchmarking and information utility-aware WSI analysis, we release MUFASA, along with our precomputed masks and overlays for CAMELYON16 and multiple TCGA cohorts. 

\section{Results}

\subsection{Tasks and datasets}
Experiments were conducted on four WSI-level clinically relevant tasks to evaluate the robustness of preprocessing across heterogeneous cohorts. Diagnostic classification included tumor-versus-no-tumor classification on CAMELYON16 (breast cancer metastasis detection in axillary lymph nodes: 270 training and 129 test cases) and binary non-small cell lung cancer (NSCLC) subtyping on TCGA, comprising lung adenocarcinoma (LUAD, 380 cases) and squamous cell carcinoma (LUSC, 473 cases). Molecular biomarker prediction evaluated MSS/MSI status in colorectal cancer (CRC) using both an in-house Stanford CRC (755 cases) and TCGA-CRC (303 cases, comprising colon (COAD) and rectal adenocarcinoma (READ)) datasets. Prognostic modeling was evaluated on TCGA-STAD (364 gastric cancer cases) and TCGA-LUAD (376 cases). Collectively, these datasets span multiple organs, institutions, scanners, H\&E staining profiles, artifact types, and image resolutions, providing a realistic setting for testing the stability of preprocessing across heterogeneous WSI cohorts.

\subsection{Baseline comparison, tiling and feature extraction}
MUFASA was compared against three widely used state-of-the-art CPath preprocessing baselines: CLAM \cite{16}, Trident \cite{17}, and Histolab \cite{18}. For each method, WSI-level tissue masks were generated and used for downstream tile extraction. For MUFASA, tiles from Set1 were used for the downstream tasks, while Set2 and Set3 tiles were excluded due to their limited task-relevant morphologic content. To maintain consistent physical tissue coverage across slides, tiles were extracted using a resolution-normalized tiling strategy based on each slide’s native microns-per-pixel (MPP). WSI with MPP$\approx$0.25 were tiled at $512 \times 512$ pixels and subsequently downsampled to $256 \times 256$ for feature extraction. WSI with MPP$\approx$0.5 were tiled directly at $256 \times 256$, resulting in comparable tissue area per tile across datasets. For tumor diagnosis, subtyping, and MSS/MSI prediction, tile embeddings were extracted using a ResNet-50 \cite{45} encoder to minimize model-specific inductive bias and isolate the contribution of preprocessing. For survival prediction, embeddings were generated using UNI \cite{46}, a histopathologic foundation model whose richer domain-specific representations are better suited to capturing subtle prognostic signals.

\subsection{MIL benchmarking for classification and biomarker status prediction}
For the three classification tasks, 13 state-of-the-art MIL models representing diverse architectural paradigms were evaluated. These included models based on pooling (Mean-pooling), attention (DeepAttnMISL \cite{47}, CLAM-MB \cite{16}, ACMIL \cite{48}, Additive MIL \cite{49}, DTFD-MIL \cite{50}), transformers/sequences (TransMIL \cite{51}, TDA-MIL \cite{52}), dual streams (DSMIL \cite{53}), graphs/relations (ILRA-MIL \cite{54}, DGR-MIL \cite{55}), hard-instance mining (MHIM-MIL \cite{56}), and state-space modeling (MambaMIL \cite{57}). All models were used in their original form without architectural modification. Training was performed using Adam optimization with a learning rate of $2 \times 10^{-4}$, weight decay of $2 \times 10^{-5}$, and dropout of 0.2, for a maximum of 200 epochs, with early stopping when validation Macro-AUC (area under the curve) failed to improve for 40 epochs. CAMELYON16 followed the official fixed training and test split, and each model was evaluated across five random seeds. The tumor subtyping and MSS/MSI experiments employed fivefold cross-validation (4:1 split). Performance metrics included accuracy (ACC), balanced accuracy (BACC), Macro-AUC, and Macro-F1, reported as mean$\pm$standard deviation (s.d.) across runs or folds. To accurately represent model performance, bootstrap-based 95\% confidence intervals were computed for Macro-AUC to estimate performance variability under each preprocessing method. To test whether MUFASA produced consistent improvements across MIL models and evaluation metrics (ACC, BACC, Macro-AUC and Macro-F1), paired Wilcoxon signed-rank tests were applied between MUFASA and each baseline (CLAM, Histolab, and Trident). This assessed whether gains were systematic across architectures rather than driven by isolated models or random variation. To further examine the impact of preprocessing on representation quality, latent feature distributions were analyzed to assess class separability, and attention heatmaps were generated to evaluate whether preprocessing reduced artifact-associated attribution and improved focus on biologically meaningful tissue regions.

\begin{figure*}[t!] 
	\centering 
	\includegraphics[width=0.98\textwidth]{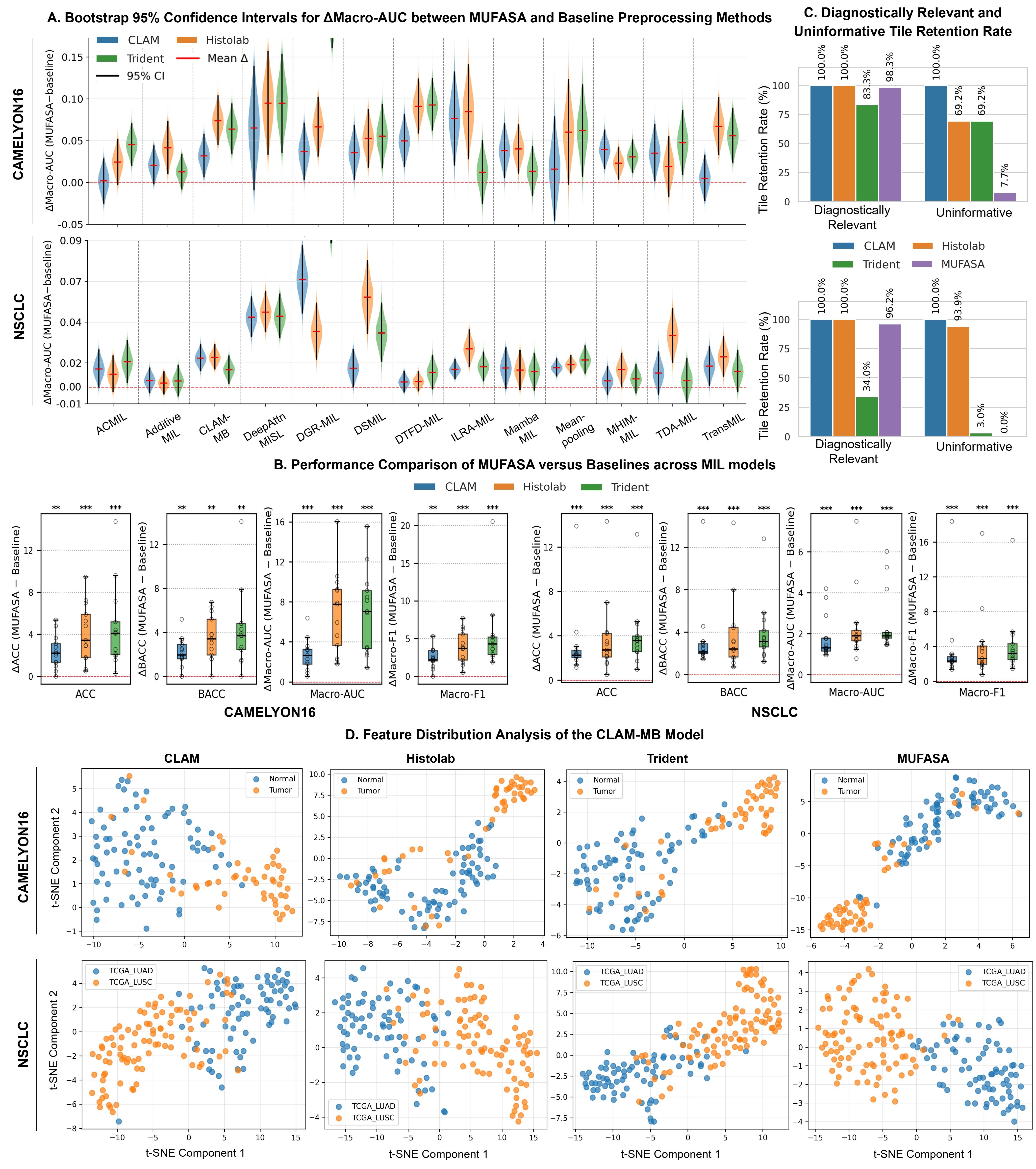} 
	\caption{\footnotesize \textrm{\textbf{MUFASA improves diagnostic robustness and reduces artifact contamination across tumor diagnosis and subtyping tasks.} \textbf{A}. Bootstrap-based confidence intervals for paired Macro-AUC differences between MUFASA and baseline preprocessing methods. For each MIL model on CAMELYON16 and NSCLC, out-of-fold predictions from five folds/runs were pooled, and stratified bootstrapping was performed by resampling the same slide indices for MUFASA and the corresponding baseline (CLAM, Histolab or Trident). $\Delta$Macro$-$AUC was calculated as MUFASA$−$baseline Macro$-$AUC over 20000 bootstrap iterations, and 95\% confidence intervals were defined using the 2.5th and 97.5th percentiles. Positive values indicate improved Macro$-$AUC with MUFASA compared to baseline preprocessing methods. Outlier paired-difference distributions are annotated with their mean $\Delta$Macro$-$AUC and 95\% confidence interval in the figure.    
    \textbf{B}. Statistical significance was assessed using a paired Wilcoxon signed-rank two-tailed test across $N=13$ MIL models. Boxplots show the distribution of model-level performance differences ($\Delta$=MUFASA$-$baseline) across evaluation metrics (ACC, BACC, Macro-AUC, and Macro-F1). Each point represents one MIL model averaged across runs/folds. The red dashed line indicates zero difference. Positive shifts across all metrics indicate improved downstream performance with MUFASA. Statistical significance is denoted as $\ast \ast \ast p<0.001$, $\ast \ast p<0.01$, $\ast p<0.05$, and $ns$, not significant.     
    \textbf{C}. Expert-annotated tile retention analysis quantifying the proportions of diagnostically relevant and uninformative tiles retained by each preprocessing method. Permissive pipelines preferentially retain more uninformative tiles, whereas aggressive filtering reduces diagnostically relevant tiles.    
    \textbf{D}. t-SNE visualization of latent feature distributions from the CLAM-MB model on CAMELYON16 and NSCLC, showing that MUFASA yields tighter within-class structures and reduced inter-class overlap relative to baseline preprocessing methods.  }}
	\label{fig:3}       
\end{figure*} 

\subsection{Impact on WSI classification }
\subsubsection{Tumor diagnosis and subtyping}
On both tumor-versus-no-tumor classification on CAMELYON16 and tumor subtyping (LUAD versus LUSC) on the TCGA NSCLC cohort, MUFASA consistently outperformed CLAM, Histolab, and Trident across the 13 MIL models. Observed ACC, BACC, Macro-AUC, and Macro-F1 scores are summarized in Supplementary Tables~\ref{tab:1} and~\ref{tab:2} (Supplementary Table ~\ref{tab:6} reports MIL model performance when all tiles from MUFASA Sets 1-3 were used). Mean performance across MIL models was computed by averaging the mean metric for each model over five runs or folds, without weighting across models. The performance gains were not confined to a single model family, indicating that the benefit arose primarily from preprocessing rather than architecture-specific optimization. On CAMELYON16, MUFASA achieved the highest average ACC across all MIL models, at 78.16, improving over CLAM (75.82) by +2.34 points (2.9\% relative), Histolab (74.04) by +4.12 points (5.2\%), and Trident (73.55) by +4.61 points (5.8\%). Comparable gains were observed for BACC, Macro-AUC, and Macro-F1. Relative to CLAM, Histolab, and Trident, MUFASA improved BACC by +2.11, +3.52, and +4.23 points, Macro-AUC by +2.72, +7.02, and +7.01 points, and Macro-F1 by +2.41, +3.86, and +5.47 points, respectively. On NSCLC, MUFASA again improved average performance across models, achieving an ACC of 87.18, compared to CLAM (83.96; +3.22 points, 3.6\% relative), Histolab (83.36; +3.82 points, 4.3\%), and Trident (83.15; +4.03 points, 4.6\%). Improvements in Macro-AUC were smaller in absolute magnitude but remained consistent (+1.84, +2.37, and +2.57 points, respectively), while Macro-F1 increased by 3.71–4.28 points across baselines. These gains are clinically meaningful, as both tumor diagnosis and subtyping rely on stable recognition of subtle morphological cues, which can be compromised by retained artifacts or excessive removal of informative tiles. 

Across CAMELYON16 and NSCLC, MUFASA showed consistent improvements over almost all baseline preprocessing methods for tumor classification and subtyping. Because model selection was driven by Macro-AUC, we computed the 95\% confidence intervals (CIs) for the $\Delta$Macro-AUC between MUFASA and each baseline using a stratified, paired bootstrap at the slide level (Fig.~\ref{fig:3}A). For NSCLC, we first obtained out-of-fold predictions from 5-fold cross-validation, so that each slide contributed exactly one prediction per model from a fold in which it was held out. We then jointly resampled slide indices with replacement within each class (stratified bootstrap), preserving pairing between our model and the baseline for the same resampled slides, and recomputed $\Delta$Macro-AUC for each bootstrap sample (20,000 iterations). The 2.5th and 97.5th percentiles of the bootstrap distribution were used to form the 95\% CI. Across all 13 MIL models, MUFASA produced positive mean $\Delta$Macro-AUC values for every comparison on both CAMELYON16 and NSCLC. On CAMELYON16, the mean $\Delta$Macro-AUC across models was +0.035 versus CLAM, +0.057 versus Histolab and +0.061 versus Trident. The largest improvements were observed for DGR-MIL, particularly versus Trident (+0.210), while smaller gains were observed for ACMIL and Additive MIL, where some CIs approached or crossed zero. On NSCLC, the corresponding mean gains were +0.017, +0.021 and +0.024, respectively. DGR-MIL again showed the strongest improvement, especially versus Trident (+0.138), whereas Additive MIL showed smaller gains (+0.003–0.004 across baselines). The magnitude of improvement varied by architecture, but no single model family exclusively accounted for the benefit. Larger gains were observed in graph and attention-based models, whereas hard-instance mining, state-space and some transformer-based models showed smaller but still positive mean effects. Confidence intervals crossing zero in some models likely reflects fold-level variability and baseline-dependent model stability rather than the absence of benefit, as the direction of mean $\Delta$Macro-AUC remained positive across all model-baseline comparisons. In addition, some baseline-preprocessed datasets might have improved performance for certain models, but this may partly reflect spurious correlations from retained uninformative or low-utility regions, leading to inflated prediction probabilities. Therefore, attribution patterns should be examined alongside performance metrics to determine whether model attention is directed toward biologically meaningful tissue. Overall, the results support a consistent, architecture-agnostic improvement in Macro-AUC with MUFASA. 

Paired Wilcoxon signed-rank tests were additionally used to assess whether MUFASA improved downstream performance across all 13 MIL variants and metrics (Fig.~\ref{fig:3}B). For each metric, the test was performed on paired model-level differences between MUFASA and each baseline, where each point represented one MIL model averaged across runs or cross-validation folds. Positive differences indicated better performance with MUFASA, and the zero line represented no improvement. On CAMELYON16, MUFASA significantly improved Macro-AUC over CLAM, Histolab, and Trident by +2.66, +7.78, and +7.04 points, respectively (all $p=2.44\times10^{-4}$), with consistent gains in ACC, BACC, and Macro-F1. On NSCLC, Macro-AUC gains remained significant at +1.31, +1.90, and +1.90 points, respectively (all $p=2.44\times10^{-4}$). Overall, MUFASA provided model-agnostic gains over both permissive and aggressive preprocessing strategies. 

To further examine how preprocessing influences model learning behavior on classification tasks, latent feature distributions were visualized using t-SNE to provide qualitative insights into the observed performance gains (Fig.~\ref{fig:3}D). As attention map analysis was performed using the CLAM-MB model, feature distributions were likewise derived from the CLAM-MB model trained on CAMELYON16 and NSCLC datasets preprocessed using CLAM, Histolab, Trident, and MUFASA. With MUFASA, the test-set feature space showed cleaner class separation than with CLAM or Histolab. With those baselines, substantial overlap remained, consistent with artifact-driven variance reducing the distinctness of the latent representations. Trident produced sharper separation in some cases; however, this came at the cost of aggressive tile removal, including those from diagnostically relevant regions, resulting in overall performance degradation. This interpretation was further supported by pathologist-based tile retention analysis, which explicitly distinguished between diagnostically relevant tissue and uninformative regions. Here, we used "diagnostically relevant" to refer to tissue regions that are informative for clinical or model decision-making, whereas uninformative regions (Fig.~\ref{fig:1}A) refer to non-tissue background, low tissue-content, and artifact-containing regions that can mislead downstream analysis. Tissue regions that are histologically valid but not expected to contribute meaningfully to the target diagnostic task were designated as neutral tissue. For the evaluation shown in Fig.~\ref{fig:3}C, 11 tumor WSI were randomly selected from the CAMELYON16 test set, and a pathologist annotated 29 regions of interest (ROI), each of size $1024\times1024$ pixels, spanning these three categories (diagnostically relevant, neutral, and uninformative). Tiles were then extracted from each ROI according to the slide resolution, using $512\times512$ pixels when MPP was $\approx$0.25 and $256\times256$ pixels when MPP was $\approx$0.5. In total, 116 tiles were obtained from the 29 ROIs, including 60 diagnostically relevant tiles, 43 neutral tissue tiles, and 13 uninformative tiles. Following the same procedure, 10 tumor WSI were randomly selected from the TCGA-LUSC cohort test set, and 36 annotated ROIs of size $1024\times1024$ pixels yielded 144 tiles, comprising 53 diagnostically relevant tiles, 25 neutral tissue tiles, and 66 uninformative tiles.  

All sampled tiles were independently reviewed by pathologists to evaluate retention of diagnostically relevant and uninformative tiles across preprocessing methods. On CAMELYON16, CLAM retained 100\% of uninformative and diagnostically relevant tiles, while Histolab retained 69.2\% of uninformative tiles and 100\% of diagnostically relevant tiles. Trident also retained 69.2\% of uninformative tiles but only 81.8\% of diagnostically relevant tiles, compared to CLAM and Histolab. In contrast, MUFASA reduced uninformative tile retention to 7.7\%, while preserving 98.3\% of diagnostically relevant tiles. In the TCGA-LUSC cohort, CLAM and Histolab retained 100\% and 93.9\% of uninformative tiles, respectively, while retaining 100\% of diagnostically relevant tiles, whereas Trident retained only 3\% of uninformative tiles but preserved only 34\% of diagnostically relevant tiles. In contrast, MUFASA removed all uninformative tiles (0\%), while retaining 96.2\% of diagnostically relevant tiles. Together, these results indicate that improved diagnostic performance with MUFASA arose not from indiscriminate filtering, but from a more favorable balance between uninformative tile exclusion and tissue preservation. This is particularly relevant to clinical deployment, where an effective preprocessing framework must suppress spurious signals without discarding the tissue compartments that underpin diagnosis.  

\begin{figure*}[H] 
	\centering 
	\includegraphics[width=0.82\textwidth, height=0.71\textheight]{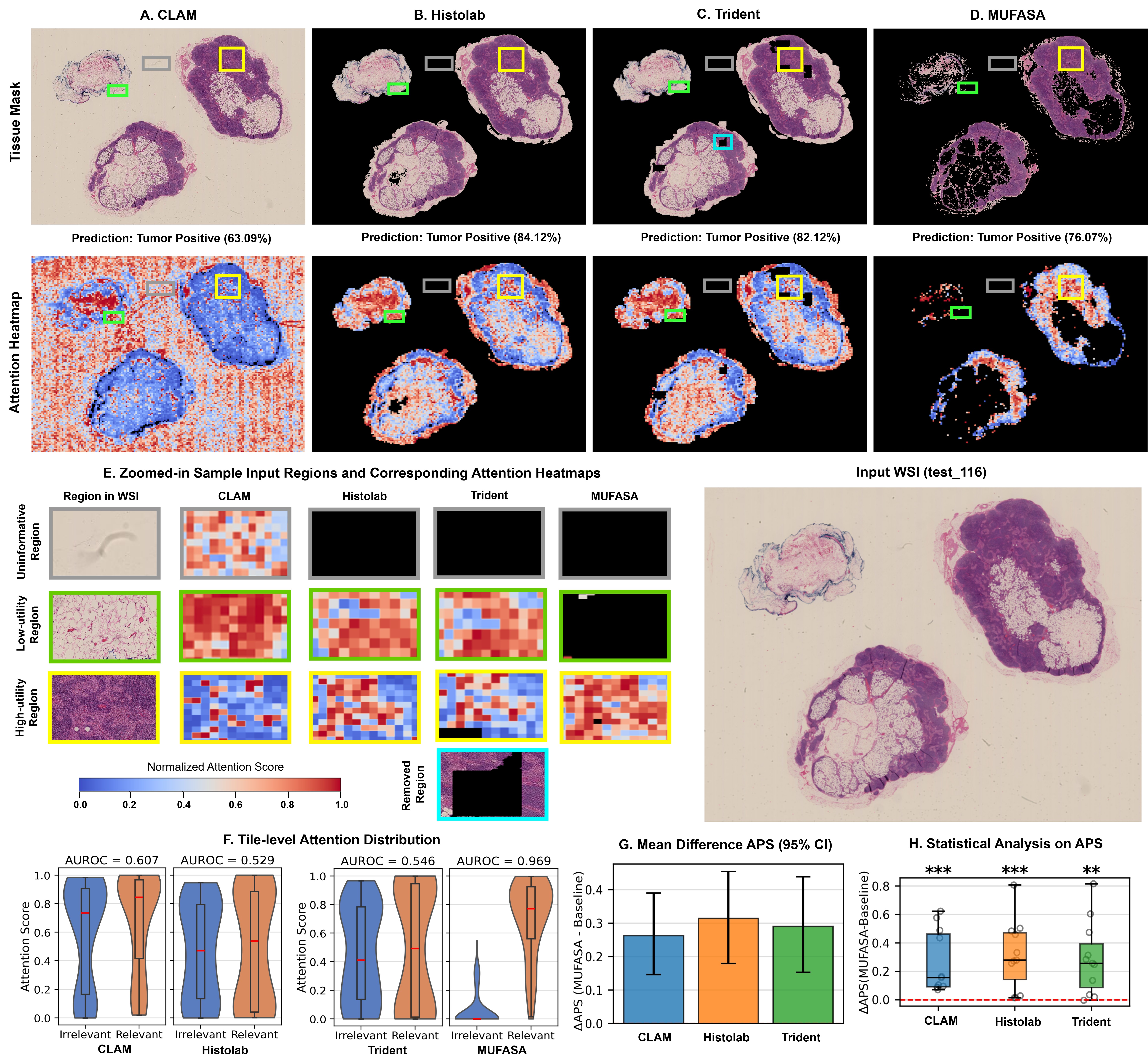} 
	\caption{\footnotesize \textrm{\textbf{Artifact-aware preprocessing improves anatomic validity of model attention in attention-based MIL.} 
    \textbf{A–D}. Tissue masks and tile-level attention heatmaps for the same representative CAMELYON16 test WSI (test\_116) generated using CLAM (A), Histolab (B), Trident (C), and MUFASA (Set1) (D). Baseline preprocessing methods show distinct failure modes: CLAM retains broad non-tissue background and artifact-prone regions; Histolab partially suppresses non-tissue background but preserves background-heavy, diagnostically irrelevant adipose-containing areas; Trident removes non-tissue regions but also excludes diagnostically relevant tissue. MUFASA yields a more selective mask enriched for tissue-dense and diagnostically relevant regions with minimal non-tissue background contamination. 
    \textbf{E}. Representative regions illustrating differences in attention localization. CLAM assigns high attention to artifact and non-tissue regions, whereas Histolab and Trident remove some non-tissue background influence, but still direct high-attention to adipose-containing or otherwise diagnostically irrelevant compartments. Predominantly black attention grids indicate regions that were removed by the preprocessor before being passed to the MIL model. Trident also under-samples tissue-rich areas. In contrast, MUFASA concentrates higher attention within diagnostically relevant tissue regions (to which the baseline methods assigned mixed or low attention), indicating improved alignment between retained tissue and model focus. 
    \textbf{F}. Tile-level attention distributions and AUROC for expert-annotated diagnostically relevant versus irrelevant (neutral tissue and uninformative) tiles from CAMELYON16. MUFASA shows the strongest separation between diagnostically relevant and irrelevant attention values. 
    \textbf{G}. Mean paired differences in APS ($\Delta$APS=MUFASA$−$baseline) computed at the slide level for CLAM, Histolab, and Trident. For each comparison, paired slide-level APS differences were calculated across $N=11$ slides and resampled using 20,000 bootstrap iterations to estimate the mean effect and 95\% confidence interval. Positive values indicate improved attention preservation under MUFASA relative to baseline preprocessing methods, summarizing the joint preference for retaining diagnostically relevant tiles, excluding uninformative tiles, and assigning low attention to diagnostically irrelevant tissue regions. 
    \textbf{H}. Performance differences in slide-level APS ($N=11$) were evaluated using two-tailed paired Wilcoxon signed-rank tests to compare MUFASA against baseline methods. Each point represents one slide. Boxplots summarize the distribution of paired differences ($\Delta$APS=MUFASA$−$baseline), with the red dashed line indicating zero difference. Positive shifts indicate improved attention preservation under MUFASA. Significance markers indicate $\ast \ast \ast p<0.001$, $\ast \ast p<0.01$, $\ast p<0.05$, and $ns$, not significant.     }}
	\label{fig:4}    
\end{figure*}  

\subsubsection{Model interpretability analysis}
Attention-based MIL variants \cite{16}, \cite{48} enable post-hoc spatial attribution and interpretability by highlighting regions assigned high or low attention, indicating which areas contributed most strongly to the prediction. We used this to assess whether preprocessing influenced not only predictive performance but also the anatomic validity of model focus. For this analysis, attention heatmaps were generated using the CLAM-MB attention-based MIL model (CLAM-MB), which assigns tile-level attention weights reflecting the relative contribution of each tile to the slide-level prediction. Fig.~\ref{fig:4} presents a representative CAMELYON16 test WSI, along with the corresponding tissue masks and tile-level attention heatmaps generated by CLAM (Fig.~\ref{fig:4}A), Histolab (Fig.~\ref{fig:4}B), Trident (Fig.~\ref{fig:4}C), and MUFASA (Fig.~\ref{fig:4}D) preprocessing. Higher attention is shown in red and lower attention in blue. This visualization enables direct inspection of whether the model concentrated on relevant tissue regions or assigned attention to uninformative regions, such as non-tissue background, pen marks, dust/debris, air bubbles, blur, tissue folds, and low-utility tissue regions (adipose-containing tiles). Attention maps therefore served as a practical readout of whether preprocessing directed the model toward biologically meaningful tissue or toward uninformative signals. 

As demonstrated in Fig.~\ref{fig:4}, clear differences were observed across preprocessing methods in both tissue masks and attention localization (Fig.~\ref{fig:4}A-D). CLAM retained substantial background and artifact-containing regions, with the resultant heatmaps frequently assigning high attention to low-utility and uninformative-regions (Fig.~\ref{fig:4}E), despite a correct tumor diagnosis. Histolab partially excluded background regions but continued to retain adipose-containing and background-heavy regions with high attention. Trident removed background-dominant regions more aggressively. However, this was accompanied by loss of diagnostically relevant tissue regions. Although Histolab- and Trident-based models produced tumor predictions exceeding 82\% probability, high-attention regions frequently localized to adipose-containing or otherwise diagnostically irrelevant compartments rather than relevant tissue regions. Trident also removed tissue-dense regions in several areas, thereby reducing the effective diagnostic content available to the model. In contrast, MUFASA retained predominantly tissue-dense and diagnostically relevant regions while minimizing uninformative regions, resulting in attention maps mostly concentrated on diagnostically relevant tissue compartments, despite a slightly lower prediction probability than that of some baseline methods. These qualitative differences were supported by quantitative analysis of tile-level attention discrimination using the expert-annotated tiles. For this analysis, the three expert-annotated categories (diagnostically relevant, neutral, and uninformative regions) were consolidated into two groups. Neutral tissue and uninformative regions were combined and considered "irrelevant", whereas diagnostically relevant tissue regions were considered "relevant". This grouping highlights that the model predominantly attends to clinically meaningful tissue regions while reducing attention to irrelevant regions in the context of tumor diagnosis. The tile-level attention distributions (Fig.~\ref{fig:4}F) show that CLAM yielded relatively high attention values for both relevant and irrelevant tiles, indicating weak discriminative focus and substantial retention of uninformative and low-utility regions. Histolab and Trident did not improve this separation relative to CLAM, assigning high and low attention values for both relevant and irrelevant tiles. By contrast, MUFASA consistently assigned higher attention to relevant tiles and lower attention to irrelevant tiles, with this pattern observed for approximately 90\% of evaluated tiles. This effect was also reflected in the attention-based AUROC (area under the receiver operating characteristic curve), which increased from 0.60 with CLAM to 0.969 with MUFASA, corresponding to an absolute gain of 0.362 and a 59.6\% relative improvement. Compared to Histolab (0.529) and Trident (0.546), MUFASA improved AUROC by 0.44 (83.2\% relative) and 0.423 (77.5\% relative).

To quantify this effect further, an attention preservation score (APS) was computed from the pathologist-verified tile annotations.  For each of $n$ annotated tiles, a relevance score $r$ was assigned according to its tile class (diagnostically relevant, neutral tissue, and uninformative) and retention status. If a diagnostically relevant tile was retained, $r$ was set to its attention score; otherwise, $r=0$. If a neutral tissue tile was retained, $r$ was set to 1−attention score; otherwise, $r=0$. If an uninformative tile was retained, $r=0$, whereas if the uninformative tile was removed during preprocessing, $r=1$. APS was then computed as the average of $r$ across all annotated tiles. Under this definition, a higher APS indicates a more desirable attention profile, characterized by retention of diagnostically relevant tiles with high attention, retention of neutral tissue tiles with low attention, and successful removal of uninformative tiles. At the slide level, MUFASA achieved consistently positive mean paired APS improvements over all baseline preprocessing methods, with $\Delta$APS values of +0.263 over CLAM, +0.313 over Histolab, and +0.286 over Trident (Fig.~\ref{fig:4}G). Bootstrap confidence intervals were estimated from slide-level paired APS differences between MUFASA and each baseline, providing uncertainty estimates for the observed attention-preservation gains. These improvements were further supported by paired Wilcoxon signed-rank tests performed on slide-level APS scores (Fig.~\ref{fig:4}H), where each data point corresponded to one slide. MUFASA showed significantly higher APS than all three baselines (p < 0.01), indicating that its attention profile was consistently more favorable across slides. Collectively, these findings indicate that MUFASA improved the clinical plausibility of model attention by reducing attribution to uninformative and low-utility regions and concentrating attention on diagnostically relevant histomorphology. 

\begin{figure*}[t!] 
	\centering 
	\includegraphics[width=0.8\textwidth]{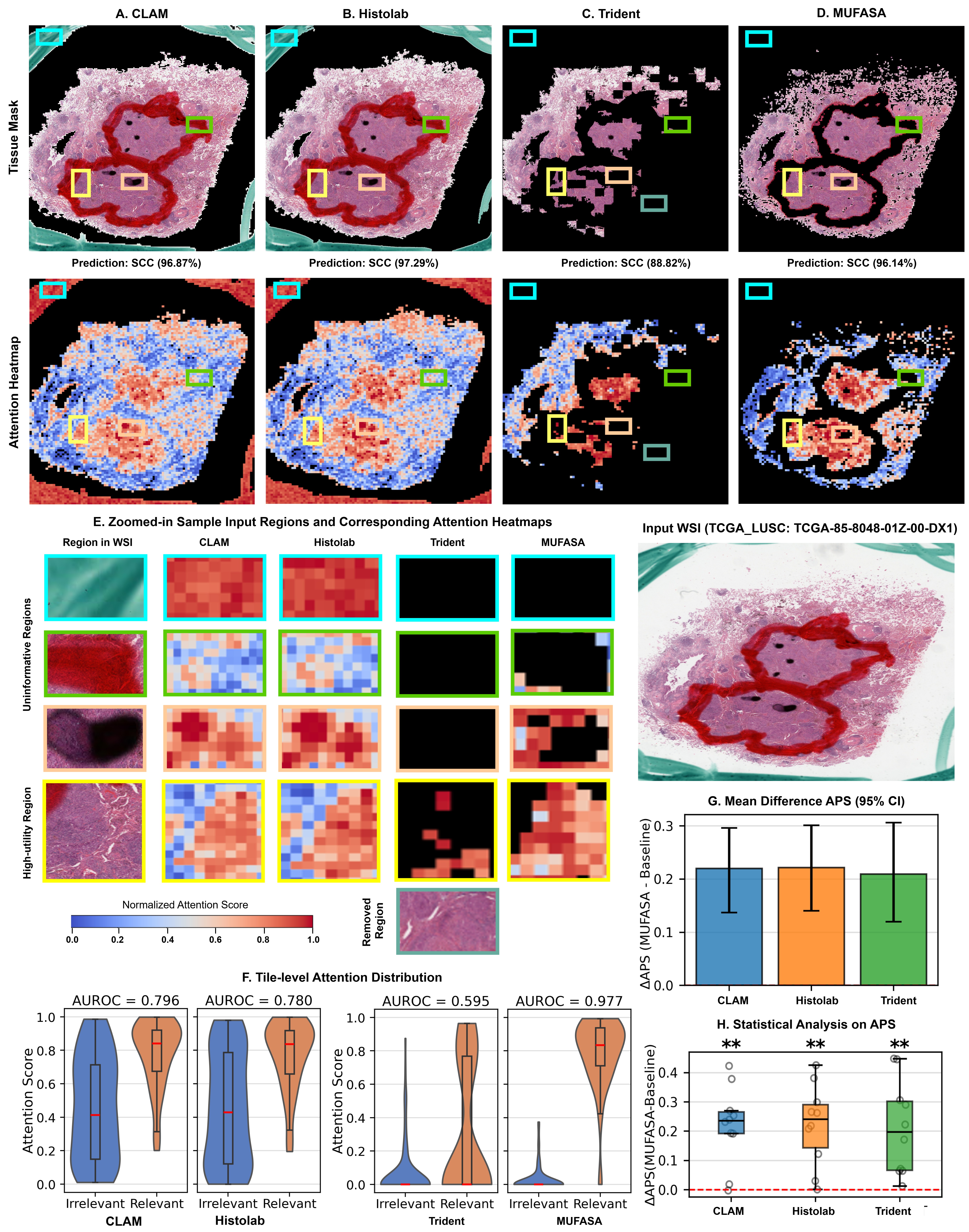} 
	\caption{\footnotesize \textrm{\textbf{Information utility-aware preprocessing improves attention localization in slides containing severe pen-mark artifact.} A representative TCGA-LUSC slide containing extensive pen-mark artifact was used to assess preprocessing robustness in an artifact-dominant setting. 
    \textbf{A-D}. Tissue masks and tile-level attention heatmaps generated using CLAM (A), Histolab (B), Trident (C), and MUFASA (Set1) (D). CLAM and Histolab retain large pen-marked regions (green and red ink), allowing artifact-dominant areas to enter the model input. Trident removes pen marks but also eliminates substantial diagnostically relevant tissue regions. MUFASA removes pen marks while preserving tissue-dense regions, producing a more anatomically correct tissue mask. 
    \textbf{E}. Representative regions with corresponding attention patterns. Despite tumor (squamous cell carcinoma (SCC)) predictions exceeding 96\% probability, CLAM and Histolab preprocessing assign high attention to pen marks and low-utility regions. Trident removes artifacts but under-samples tumor-rich tissue regions, resulting in a prediction probability of 88\%. MUFASA suppresses artifact-containing regions while concentrating attention on diagnostically relevant tissue (to which the three other preprocessing frameworks assigned only mixed or low attention) with prediction probability of 96\%. 
    \textbf{F}. Tile-level attention distributions and AUROC for expert-annotated tumor versus irrelevant (neutral tissue and uninformative) tiles from TCGA-LUSC. Consistent with CAMELYON16, MUFASA shows the strongest separation between tumor and irrelevant tile attention values. 
    \textbf{G}. Mean paired differences in APS ($\Delta$APS=MUFASA−baseline) were computed at the slide level for CLAM, Histolab, and Trident. Bootstrap-derived 95\% confidence intervals, estimated from slide-level paired differences across $N=10$ slides using 20,000 iterations, indicate consistently positive gains under MUFASA, comparable to CAMELYON16. 
    \textbf{H}. Two-tailed paired Wilcoxon signed-rank tests on slide-level APS ($N=10$). Each point represents one slide; boxplots summarize $\Delta$APS distributions. Positive shifts relative to zero indicate improved attention preservation under MUFASA. Significance markers: $\ast \ast \ast p<0.001$, $\ast \ast p<0.01$, $\ast p<0.05$, and $ns$, not significant.     }}
	\label{fig:5}       
\end{figure*} 

A second example (Fig.~\ref{fig:5}) from the TCGA-LUSC cohort demonstrates that the attention-localization behavior observed in Fig.~\ref{fig:4} is not dataset-specific. The selected WSI contained extensive pen-mark artifact, presenting a challenge for preprocessing pipelines. CLAM and Histolab retained most pen-marked regions, which subsequently attracted high attention, despite tumor predictions exceeding 96\% probability. Trident removed the pen marks but also eliminated substantial tissue-rich areas, reducing the effective diagnostic context available to the model. MUFASA selectively removed pen marks while preserving diagnostically relevant tissue. Quantitatively, MUFASA achieved higher attention discrimination, improving APS across slides by 0.21 over CLAM, 0.215 over Histolab, and 0.207 over Trident, while retaining 96.23\% of diagnostically relevant tiles and removing 100\% of uninformative tiles. These findings show that information utility-aware preprocessing improved not only predictive robustness but also the anatomic validity of model attribution in artifact-rich WSI. More broadly, the analyses demonstrate that preprocessing influences model learning capability, directing it toward relevant histomorphology rather than non-tissue background, artifacts, or over-retained low-information tissue regions.

\subsubsection{Impact on Biomarker Status Prediction (MSS/MSI) }
\begin{figure*}[t!] 
	\centering 
	\includegraphics[width=0.93\textwidth, height=0.93\textwidth]{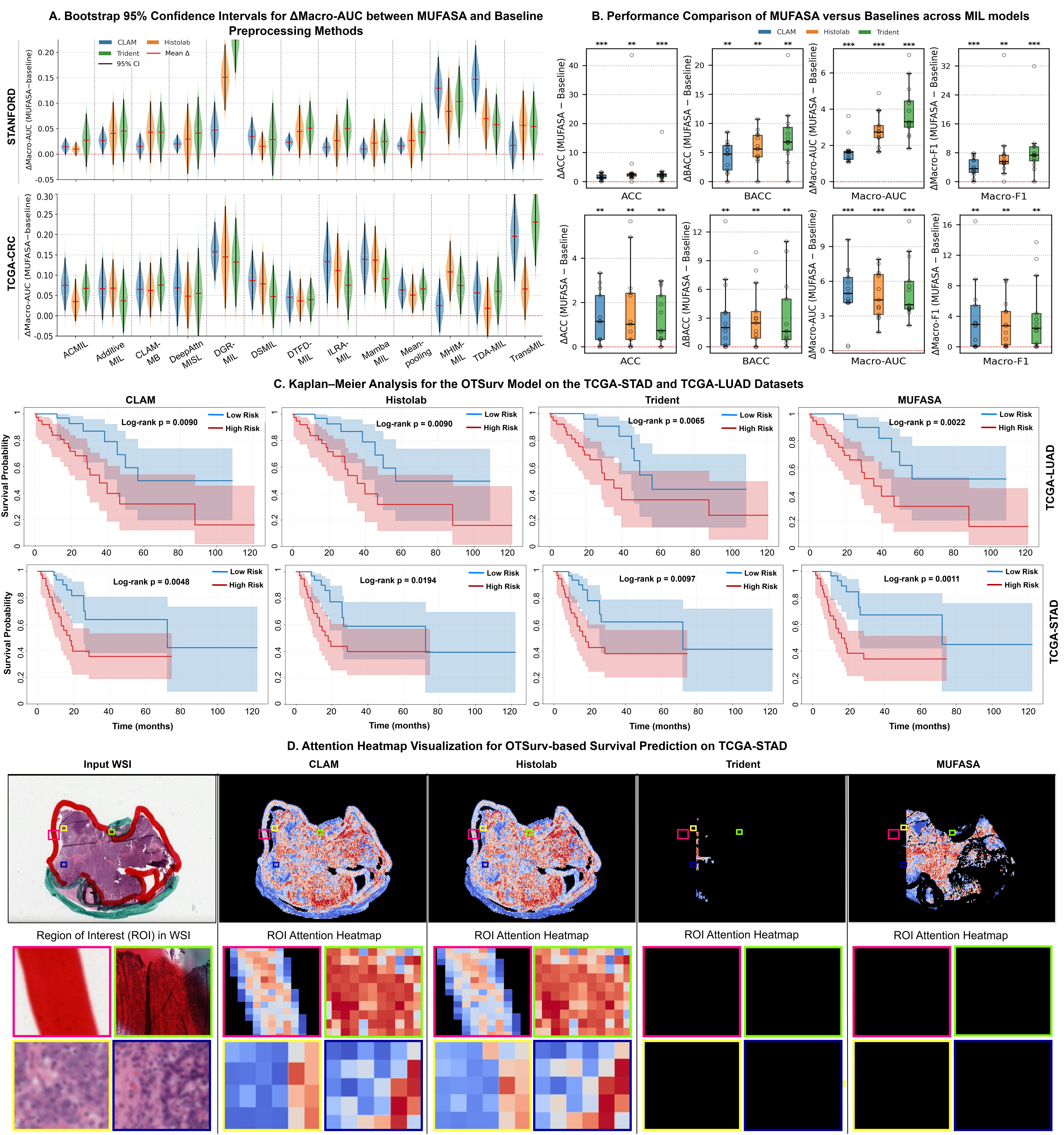} 
	\caption{\footnotesize \textrm{\textbf{Molecular biomarker prediction and survival analysis. }
    \textbf{A}. Bootstrap-based confidence intervals for paired Macro-AUC differences between MUFASA and baseline preprocessing methods for MSS/MSI prediction. For each MIL model on the Stanford CRC and TCGA-CRC cohorts, out-of-fold predictions from five folds were pooled and stratified bootstrapping was performed by resampling the same slide indices for MUFASA and the corresponding baseline (CLAM, Histolab, or Trident). $\Delta$Macro-AUC was computed as MUFASA−baseline over 20,000 bootstrap iterations, and 95\% confidence intervals were estimated using the 2.5th and 97.5th percentiles. Positive values indicate improved Macro-AUC with MUFASA. Outlier paired-difference distributions are annotated with their mean $\Delta$Macro-AUC and confidence interval. 
    \textbf{B}. Statistical comparison of preprocessing methods using two-tailed paired Wilcoxon signed-rank tests across 13 MIL models for both datasets. Boxplots show the distribution of model-level performance differences ($\Delta$=MUFASA−baseline) across evaluation metrics (ACC, BACC, Macro-AUC, and Macro-F1). Each point represents one MIL model averaged across folds. The red dashed line indicates zero difference. Positive paired differences indicate improved performance with MUFASA preprocessing. On the Stanford cohort, MUFASA yields median improvements of 1.43–2.34\% in ACC, 4.74–6.80\% in BACC, 1.62–3.30\% in macro-AUC, and 3.60–7.32\% in macro-F1 relative to baseline pipelines. On TCGA-CRC, improvements include 0.73–1.14\% in ACC, 1.62–2.48\% in BACC, 3.96–4.95\% in macro- AUC, and 2.45–2.92\% in macro-F1. All comparisons show statistically significant differences ($p<0.01–0.001$, $\ast \ast \ast p<0.001$, $\ast \ast p<0.01$, $\ast p<0.05$, and $ns$, not significant), indicating that artifact-aware preprocessing improves MSI prediction robustness across independent cohorts. 
    \textbf{C}. Kaplan–Meier curves for TCGA-LUAD (top) and TCGA-STAD (bottom), generated using OTSurv survival model with median predicted risk to stratify patients into high- and low-risk groups for each preprocessing strategy. Log-rank p-values are shown in each panel; shaded bands denote confidence intervals. 
    \textbf{D}. Representative TCGA-STAD WSI with attention overlays from a prognostic MIL model after preprocessing with CLAM, Histolab, Trident, and MUFASA. Manually selected ROIs include pen marks, pen marks overlying tissue folds, and blurry regions. The corresponding ROI-level attention grids are shown to the right. Predominantly black overlays and near-zero ROI attention grids indicate that uninformative and low-utility tiles were largely removed before passing to the MIL model, reducing the number of confounding instances input into the prognostic model.     }}
	\label{fig:6}       
\end{figure*}

Molecular biomarker status prediction was evaluated using microsatellite instability (MSS/MSI) classification on two independent cohorts: an in-house Stanford CRC and a TCGA-CRC cohort. The objective was to determine whether information utility-aware preprocessing improves biomarker status prediction by guiding model learning toward biologically relevant tissue regions rather than spurious visual signals. Across both cohorts and multiple MIL models, MUFASA consistently improved predictive performance relative to CLAM, Histolab, and Trident. Observed ACC, BACC, Macro-AUC, and Macro-F1 for all models are summarized in Supplementary Tables~\ref{tab:3} and~\ref{tab:4}. MUFASA consistently achieved a higher mean Macro-AUC compared to CLAM, Histolab, and Trident across MIL models on the Stanford cohort, indicating robust, model-agnostic performance improvements (Fig.~\ref{fig:6}A). Macro-AUC uncertainty was assessed using stratified bootstrapping of paired out-of-fold predictions, following the same procedure described previously. For each MIL model, the same slide indices were resampled for MUFASA and each baseline preprocessing method, $\Delta$Macro-AUC was computed over 20,000 iterations, and the 2.5th–97.5th percentiles defined the 95\% confidence interval. On the Stanford CRC cohort, MUFASA showed positive mean $\Delta$Macro-AUC across all models and baselines, with average gains of +0.040 over CLAM, +0.048 over Histolab and +0.062 over Trident. The strongest improvements were observed for DGR-MIL versus Trident (+0.231, 95\% CI 0.160–0.301), TDA-MIL versus CLAM (+0.147, 0.083–0.214), and MHIM-MIL versus CLAM (+0.130, 0.071\\–0.191). Smaller effects were observed for ACMIL, MambaMIL and ILRA-MIL, where some intervals crossed zero, indicating fold-level variability rather than consistent loss. On TCGA-CRC, the effect was larger overall, with average gains of +0.091 over CLAM, +0.074 over Histolab and +0.081 over Trident. TransMIL showed the largest gains versus Trident (+0.231, 0.123–0.337) and CLAM (+0.196, 0.086–0.305), while DGR-MIL and MambaMIL also showed strong positive 95\% CIs. As with tumor classification and subtyping, the gains were distributed across model families, rather than being restricted to one architecture, supporting MUFASA’s role as a preprocessing-level improvement. In addition, all paired differences were statistically significant, as determined by the paired Wilcoxon signed-rank test (Fig.~\ref{fig:6}B). A similar pattern was observed for TCGA-CRC, indicating that the effect generalized across independent cohorts. MUFASA yielded statistically significant median ACC gains of 0.73–1.14\%, BACC gains of 1.62–2.48\%, Macro-F1 gains of 2.45–2.92\%, and Macro-AUC gains of 3.96–4.95\% relative to baselines. Consistent with the results of the interpretability analyses from the tumor classification tasks, we found that MUFASA shifted the effective training distribution toward biologically relevant morphology by reducing artifact-dominant tiles and preserving tissue-rich regions. This improved the ability of MIL models to capture subtle histomorphologic correlates of MSI status, thereby increasing both predictive performance and robustness across cohorts.

\subsection{Impact on Survival Analysis}
For survival analysis, 13 MIL baselines were evaluated, including pooling methods (Mean-pooling, Max-pooling), attention-based approaches (CLAM-MB \cite{16}, DSMIL \cite{53}, DTFD-MIL \cite{50}), sequence-based models (TransMIL \cite{51}, MambaMIL \cite{57}), graph-based methods (DeepAttnMISL \cite{47}, IBMIL \cite{58}), and survival-oriented architectures (OTSurv \cite{59}, ILRA-MIL \cite{54}, PANTHER \cite{60}). Models were optimized using AdamW (learning rate $2 \times 10^{-4}$) and trained up to 200 epochs with early stopping based on validation C-index. Fivefold cross-validation (4:1) was applied and performance reported as held-out concordance index (C-index; mean$\pm$s.d.). Kaplan–Meier survival curves were used to visualize risk stratification across patient groups. We evaluated whether improvements in instance (tile) quality translated into more reliable prognostic modeling in gastric (TCGA-STAD) and lung (TCGA-LUAD) adenocarcinoma. Across all MIL models, MUFASA consistently outperformed CLAM, Histolab, and Trident. On TCGA-LUAD, MUFASA improved C-index by +0.016 to +0.039\% relative to CLAM across models. On TCGA-STAD, the corresponding gains ranged from +0.014 to +0.038\%. Improved prognostication with MUFASA preprocessing was not restricted to a single model or cohort, but was reproducible across diverse survival-oriented MIL formulations. These improvements are particularly meaningful because survival prediction is highly sensitive to weakly informative or confounding instances. Unlike tumor classification, prognostic signals are often sparse, spatially heterogeneous, and more vulnerable to dilution by retained artifacts or background-heavy tiles. Kaplan–Meier analysis curves were generated using the best performing survival model, OTSurv, to provide an orthogonal assessment of the clinical relevance of these improvements (Fig.~\ref{fig:6}C). Patients were stratified into high- and low-risk groups using the median predicted risk for each preprocessing strategy. Under this identical decision rule, MUFASA yielded the strongest and most consistent separation of survival trajectories in both cohorts. On TCGA-STAD, MUFASA achieved the lowest log-rank p-value (p=0.0011), compared with CLAM (p=0.0048), Histolab (p=0.0194), and Trident (p=0.0097). On TCGA-LUAD, MUFASA again produced the strongest separation (p=0.0022), improving over CLAM (p=0.0090), Histolab (p=0.0090), and Trident (p=0.0065). Because median-risk splits are coarse and do not directly evaluate calibration, these results are best interpreted comparatively. Nevertheless, under the same thresholding rule, the stronger and more sustained separation achieved with MUFASA indicates that risk scores captured clinically meaningful variation more effectively than those derived from baseline preprocessing pipelines.

\begin{figure*}[t!] 
	\centering 
	\includegraphics[width=0.95\textwidth]{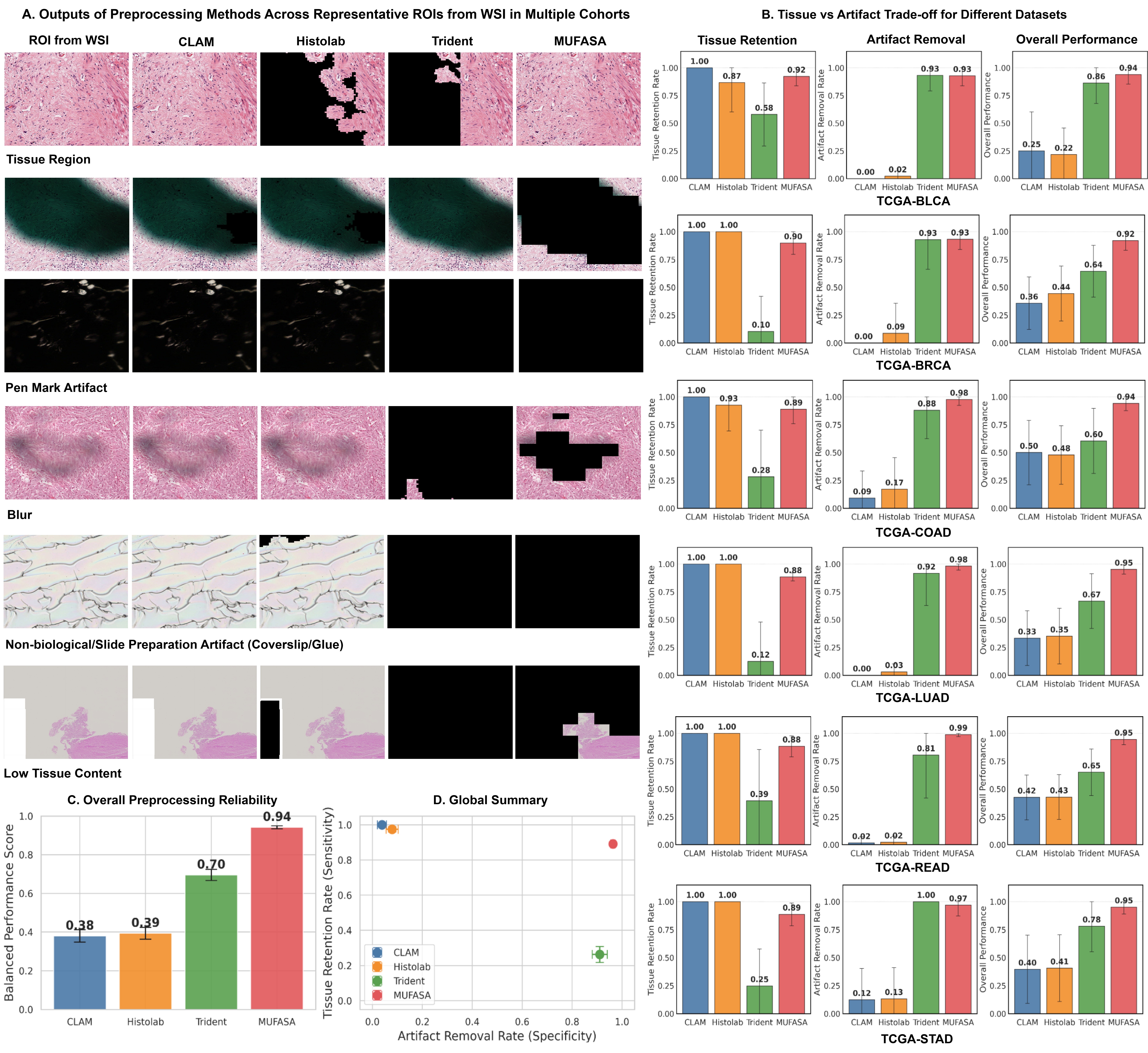} 
	\caption{\footnotesize \textrm{\textbf{Regional stress-test reveals distinct preprocessing operating points under common pathology failure modes. }
    \textbf{A}. Representative 2048×2048-pixel ROIs manually selected by a pathologist to illustrate five recurrent preprocessing challenge categories. For each ROI, outputs generated by CLAM, Histolab, Trident and MUFASA are shown using identical display settings. Black regions indicate regions excluded during preprocessing. These examples illustrate the failure modes observed throughout the study: permissive artifact retention (CLAM, Histolab) and over-aggressive tissue removal (Trident). In contrast, MUFASA demonstrates selective artifact suppression while preserving tissue continuity. 
    \textbf{B}. Cohort-level stress-test across six TCGA cohorts (BLCA, BRCA, COAD, LUAD, READ and STAD) using 72 ROIs (2 ROIs per WSI × 6 WSI per cohort × 6 cohorts). Bar graphs show the tissue retention rate (\%), artifact removal rate (\%), and their mean as balanced performance (\%) for each preprocessing pipeline. 
    \textbf{C}. Aggregate balanced preprocessing reliability across cohorts. MUFASA achieved the strongest overall score (94\%), compared to 70\% for Trident and 38–39\% for CLAM and Histolab. 
    \textbf{D}. Global summary across the sensitivity–specificity space. CLAM and Histolab clustered in the high-sensitivity/low-specificity region, Trident in the high-specificity/low-sensitivity region, and MUFASA in the favorable high-sensitivity/high-specificity region, indicating improved handling of the tissue-versus-artifact trade-off across heterogeneous pathology failure modes.    }}
	\label{fig:7}       
\end{figure*} 

To elucidate possible mechanisms underpinning these improvements, we performed attention heatmap analysis on a representative TCGA-STAD slide containing multiple artifact types (Fig.~\ref{fig:6}D). Four ROI representing tile-quality regimes relevant to preprocessing were selected: a pen-mark artifact overlying non-tissue background, multiple pen-mark artifacts overlying tissue and tissue-fold artifact, and two blurry regions with limited cellular definition. With CLAM and Histolab, attention maps remained dense across broad tissue regions, and the corresponding ROI-level attention grids showed non-trivial attention within these ROIs. This pattern indicates that large numbers of confounding tiles were passed into MIL aggregation, forcing the prognostic model to process instances unlikely to contribute meaningfully to outcome prediction. In contrast, Trident and MUFASA removed these and artifact-prone regions before MIL aggregation. However, there was a key distinction between the two approaches. Trident achieved this result through aggressive filtering that also removed portions of informative tissue, consistent with reduced tissue recall. MUFASA, in contrast, more selectively suppressed uninformative regions while preserving informative tissue regions on the slide. This balance between uninformative region suppression and tissue preservation provides a plausible explanation for the consistent gains in C-index and the stronger Kaplan–Meier separation observed across both cohorts. More broadly, these results suggest that preprocessing should be treated as a critically important component of deep learning-based prognostic pipelines and evaluated with the same rigor as MIL architecture.

\subsection{Pathologist evaluation and quality control via regional stress-test on $2048\times2048$ ROIs}
Slide-level aggregation in MIL models can obscure preprocessing failure modes. To obtain a more granular assessment, a regional stress-test was performed using fixed-size $2048\times2048$ -pixel ROIs manually selected by a pathologist to represent common preprocessing challenges. Five predefined categories were considered: artifact-free tissue, ink or pen-mark artifacts, scanning artifacts (blur), high-frequency non-biological contaminants (dust, coverslip scratches, glue), and tissue fragmentation or low tissue content (Fig.~\ref{fig:7}A). For each cohort, the pathologist selected two representative ROIs per WSI across six slides, and this procedure was repeated across the six TCGA cohorts (BLCA, BRCA, COAD, LUAD, READ and STAD), yielding 72 ROIs in total. Each ROI was partitioned into an 8×8 grid of 256×256 tiles (64 tiles per ROI) at the highest magnification based on MPP. Identical ROI coordinates were then evaluated across the masks generated by CLAM, Histolab, Trident and MUFASA, enabling direct comparison of preprocessing behavior independent of ROI selection. 

Qualitatively, the retained-tile masks revealed distinct failure modes. CLAM and Histolab frequently preserved substantial portions of pen-marked regions, scanning artifacts and contaminant-dense regions, indicating incomplete artifact suppression. These observations were consistent with the earlier tile-retention analyses on CAMELYON16 and TCGA-LUSC, where permissive pipelines retained 93.9–100.0\% of annotated uninformative tiles in the NSCLC cohort. Trident removed many of these artifact regions more effectively but often eliminated large fractions of relevant tissue, consistent with the earlier observation that aggressive filtering reduced diagnostically relevant tile retention to 34\% in TCGA-LUSC and 83.3\% in CAMELYON16. In contrast, MUFASA showed the most balanced behavior, removing artifact-heavy regions while preserving tissue-rich areas, including fragmented or low-density tissue regions that are prone to over-filtering.

This trade-off was quantified using the tissue extraction rate (sensitivity), artifact removal rate (specificity with respect to artifacts), and their average, as a balanced performance score (Fig.~\ref{fig:7}B). Across cohorts, CLAM and Histolab showed near-maximal tissue retention (often close to 100\%), but artifact removal was limited to only 0–17\%, yielding an overall balanced performance of 33–48\%. Trident achieved much higher artifact removal (81–100\%), but tissue retention decreased to 10–58\%, resulting in a moderate overall performance of 60–86\%. In contrast, MUFASA maintained 93–99\% artifact removal while preserving 88–92\% of tissue, producing the strongest balanced performance across all cohorts, at 92–95\%. At the aggregate level, MUFASA achieved the highest overall preprocessing reliability (94\%), compared to 70\% for Trident and 38–39\% for CLAM and Histolab (Fig.~\ref{fig:7}C). The sensitivity–specificity summary shows that MUFASA occupies a favorable operating range characterized by both high tissue retention and high artifact suppression, whereas CLAM and Histolab favor sensitivity at the expense of specificity, and Trident favors specificity at the expense of tissue preservation (Fig.~\ref{fig:7}D). These results demonstrate that MUFASA reduces artifact leakage without collapsing tissue coverage, which is the desired preprocessing regime for downstream MIL training and inference.

\section{Discussion}
This study demonstrates that WSI preprocessing is not simply a routine upstream operation in CPath, but a critical determinant of downstream AI model reliability, interpretability, and efficiency. Across diagnostic, tumor classification, molecular biomarker status prediction, and survival modeling tasks, MUFASA consistently improved downstream MIL performance by selectively removing uninformative and low-utility regions while preserving diagnostically relevant tissue regions. These findings suggest that, in weakly supervised learning, improved performance may arise not only from adding informative regions, but also from strategically removing uninformative regions that confound downstream representation learning and attention aggregation. In addition, slide-level accuracy alone is insufficient to assess preprocessing quality, because models may achieve correct predictions while relying on anatomically implausible regions. Correct slide-level predictions may mask significant flaws in the underlying model when the retention of tiles containing uninformative regions, such as pen marks, debris, air bubbles, blur, tissue folds, scanner stripe artifacts, and stain-related artifacts, results in anatomically implausible (i.e., spurious) reasoning by the prediction model. In such cases, the apparently successful predictions may be dependent on unstable correlations that are unlikely to generalize across heterogeneous cohorts. In certain cases, MUFASA did not produce the greatest quantitative improvement over competing preprocessing pipelines. However, detailed interpretability analyses revealed that, in these cases, the baseline methods achieved comparable or higher prediction confidence while assigning substantial attention to artifact-dominant regions. In contrast, MUFASA consistently concentrated attention on biologically meaningful tissue compartments, showing that improvements in attribution reliability may occur even when metric-level performance differences are modest. These findings highlight that both preprocessing quality and downstream predictive model quality should be evaluated based not only on predictive performance but on the anatomic validity of model reasoning.  

A key finding of our study is that existing preprocessing pipelines occupy distinct, suboptimal operating points along the artifact-removal versus information-retention continuum (Fig.~\ref{fig:7}D). Permissive pipelines retain large fractions of artifact-containing  (information-poor) tiles, allowing confounding instances to enter MIL aggregation and influence attention. More aggressive pipelines suppress many of these artifacts, often at the cost of discarding relevant tissue regions. MUFASA consistently achieves a more favorable balance: strong artifact suppression without substantial loss of tissue coverage. This balance was reflected in improved Macro AUC, Macro F1, and C-index values, cleaner class separation within the latent space, anatomically credible attention maps, higher attention preservation scores, and significantly better ROI-level stress-test performance across multiple TCGA cohorts. Together, these results indicate that preprocessing quality directly affects the effective instance pool seen by MIL models and therefore influences which morphologic signals are ultimately available for learning. Beyond quantitative performance, these trade-offs are particularly important in translational settings, where slide preparation, stain quality, scanning conditions and artifact burden vary across pathology laboratories and over time. A preprocessing framework intended for clinical deployment must therefore do more than remove obvious background. It must minimize artifact-driven learning while preserving the tissue compartments that contribute to diagnosis, biomarker status prediction, and prognostication. 

MUFASA was developed around commonly occurring artifacts observed in routine H\&E-stained WSI and is distinct from existing preprocessing pipelines, not only in its empirical performance, but in its architectural design and extensibility. Classical heuristic pipelines operate through fixed-intensity rules that are computationally efficient but sensitive to stain variability and scanner heterogeneity, often requiring dataset-specific tuning, which limits their adaptability to new artifact types. Widely used pipelines such as CLAM, Histolab, and Trident similarly rely on heuristic tissue masks combined with post-hoc filtering steps, which can remove coarse background but frequently leave fine-grained artifacts embedded within tissue regions. Quality-control frameworks such as HistoQC, SliDL, PathML, FAST, TIAToolbox, and GrandQC provide structured workflows for detecting blur, tissue folds, and pen marks. However, these approaches primarily rely on rule-based filtering or discrete artifact detection. Classification-based methods treat artifact detection as a discrete classification problem rather than a continuous estimation of tile utility for downstream learning. Recent segmentation-based approaches and content-aware tiling pipelines improve tissue localization using deep learning, yet these methods often require annotated training data, may struggle under domain shift, and remain focused primarily on tissue-versus-background discrimination rather than prioritization of biologically informative regions.

MUFASA differs conceptually from these approaches by formulating preprocessing as a progressive information utility-aware tile filtering problem rather than a binary segmentation task. Earlier phases remove non-tissue background and artifact (such as pen marks, dust, and debris)-dominant regions using deterministic rules, while later phases estimate reconstruction-based utility distributions that allow tiles to be stratified along a continuum of biological relevance. This design enables separation of tissue-rich tiles from low-utility (e.g., adipose) and uninformative regions. In contrast to monolithic segmentation models, MUFASA's modular structure allows additional artifact types to be incorporated without retraining the entire pipeline, improving adaptability across cohorts with heterogeneous acquisition conditions. More broadly, we argue that preprocessing should be treated as a critical component of WSI-based deep learning pipelines and evaluated with the same rigor as downstream model architectures, particularly when interpretability is expected to support clinical diagnosis and quality assurance. To facilitate further research in this area and support reproducible benchmarking, we have made the code for MUFASA and the pre-computed masks for the CAMELYON16 and the TCGA cohorts used in this study publicly available.

Our study has some limitations. First, MUFASA was evaluated primarily on H\&E-stained WSIs from TCGA, CAMELYON16, and Stanford cohorts, covering recurrent artifact-containing and low-information-content regions observed across these datasets. However, additional dataset-specific artifacts, scanner-dependent variations, institutional preparation differences, and non-H\&E stains were not systematically evaluated. The autoencoder was trained on a relatively small corpus of 40 WSIs from three sources, and although the framework showed consistent performance across the evaluated cohorts, broader external validation using community hospital archives, additional scanner models, and non-H\&E staining protocols will be important to further establish generalizability. Second, the downstream experiments primarily used Set1 (high-utility tiles), which was empirically supported for the evaluated tasks. However, the optimal utility-set selection may vary across applications. For example, low-utility but histologically valid regions, such as adipose-rich, mucin-rich, or lightly stained tissue, may be informative for specific diagnostic, prognostic, or tissue-composition tasks. Therefore, the generalizability of a Set1-only strategy across broader task types remains to be systematically tested. Third, the pathologist-guided tile retention and attention preservation analyses were based on single-reader annotations and relatively small slide subsets. Specifically, the analysis used 11 CAMELYON16 WSI and 10 TCGA-LUSC WSI, and inter-rater agreement was not assessed. Although these annotations provided expert-guided evidence that MUFASA preferentially preserves diagnostically relevant tissue and suppresses artifact-driven attention, larger multi-reader studies will be needed to further validate these interpretability-related findings. Fourth, under the present implementation, MUFASA processes a representative 3 GB WSI in approximately 2.5 minutes on an NVIDIA L40S GPU and 3–4 minutes on CPU. Although this runtime is approximately twofold higher than lightweight heuristic masking pipelines, the added cost is offset by improved tissue-mask reliability and reduced propagation of uninformative instances into downstream computational pathology models. Finally, the current evaluation focused on a defined set of MIL architectures and WSI-level tasks spanning tumor diagnosis and classification, biomarker prediction, and survival modeling. Additional architectures, clinical applications, and tile-level tasks remain to be evaluated. In particular, MUFASA’s utility-aware stratification could support tissue-composition quantification and spatial histologic modeling, but these applications were not systematically investigated here and remain important directions for future work.

In conclusion, we present MUFASA, a multi-phase artifact and information-utility aware preprocessing framework for H\&E-stained WSI that improves tissue identification in WSI for downstream computational pathology applications. Across diagnostic classification, biomarker status prediction and survival modeling, MUFASA consistently improves performance while reducing artifact-driven model attribution. By balancing artifact suppression with information preservation, the framework improves the predictive robustness and anatomic validity of model reasoning. We systematically evaluate the impact of including artifact-containing and other low informative-utility regions on downstream model performance, showing that preprocessing critically impacts model reliability and interpretability in whole-slide-based deep learning pipelines, and that robust information utility-aware preprocessing may be essential for improving the translational readiness of computational pathology systems. 

\section{Methods}
MUFASA is a four-phase preprocessing framework (Fig.~\ref{fig:2}A) designed to generate information utility-aware tissue masks from H\&E-stained WSI for downstream machine-learning applications. The framework targets the removal of uninformative regions such as those containing non-tissue background, low tissue content, and artifacts (air bubbles, dust and debris, pen marks, large tissue folds, blur, scanner and stain-related artifacts (Fig.~\ref{fig:1}A)). Rather than relying on a single heuristic or a segmentation model, MUFASA uses a progressive refinement strategy in which earlier phases suppress gross slide-level artifacts and later phases perform tile-level filtering to remove non-tissue background and persistent artifacts while preserving biologically meaningful tissue. Supplementary Fig.~\ref{fig:S1} provides a phase-wise audit trail of the filtering process, showing representative exclusions across Phases 1–3 and the selective recovery step in Phase 4. The final overlays visualize the spatial distribution of Sets 1–3 and the aggregate tissue mask. Each phase was implemented as a modular component so that the pipeline could be adapted to heterogeneous cohorts without task-specific redesign.

\begin{figure*}[t!] 
	\centering 
	\includegraphics[width=0.95\textwidth]{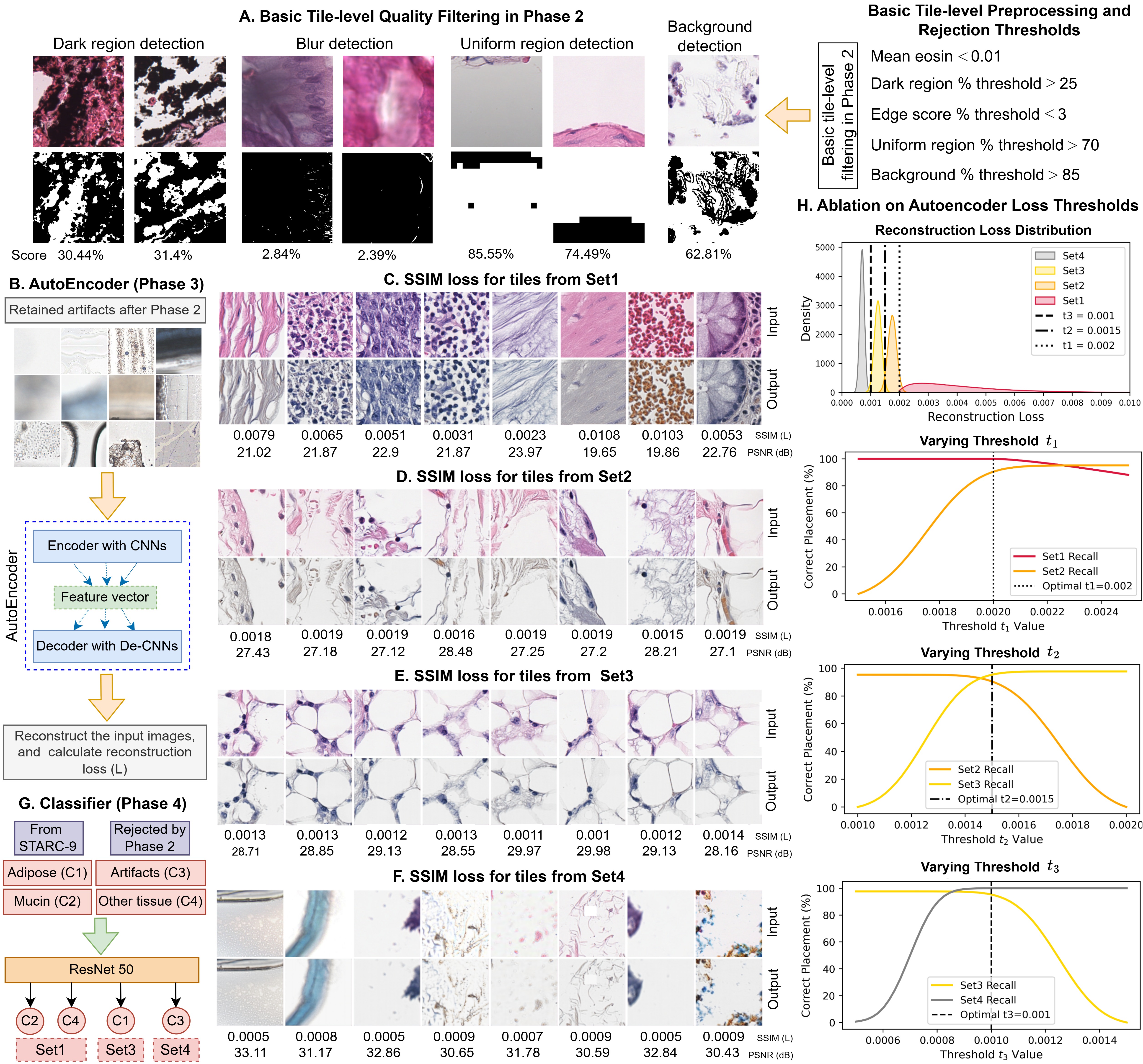} 
	\caption{\footnotesize \textrm{\textbf{Basic preprocessing, autoencoder-based tile utility filtering, and threshold ablation for MUFASA.} 
    \textbf{A}. Phase 2 basic preprocessing criteria, showing representative examples of dark region detection, blur detection, uniform region detection, and background estimation, together with the corresponding filtering scores and selected thresholds. 
    \textbf{B}. Autoencoder training module used in Phase 3, illustrating representative training inputs drawn from the uninformative (Set4) category. 
    \textbf{C–F}. Representative input tiles and corresponding autoencoder reconstructions for the four utility-aware output categories, together with reconstruction loss, structural similarity index (SSIM) and peak signal-to-noise ratio (PSNR). C. Set1: tissue-dominant tiles with high reconstruction loss. 
    \textbf{D}. Set2: partially tissue-containing tiles with intermediate reconstruction loss. 
    \textbf{E}. Set3: tiles representing adipose-containing or faintly stained tissue compartments. 
    \textbf{F}. Set4 (uninformative) tiles characterized by low reconstruction loss, corresponding to background-dominant or residual artifact-containing regions. 
    \textbf{G}. Phase 4 optional recovery module, in which tiles rejected by Phase 2 are re-evaluated by a lightweight four-class classifier to restore biologically pertinent tissue types, such as adipose or mucin, while excluding persistent artifact tiles. 
    \textbf{H}. Threshold ablation for reconstruction-loss-based utility stratification. Loss distributions are shown for four tile categories: Set1, Set2, Set3, and Set4. Vertical dashed lines indicate the selected thresholds ($t_1$=0.002, $t_2$=0.0015, $t_3$=0.001). The curves illustrate the trade-offs between adjacent categories as thresholds are shifted, and the selected operating points were chosen using Optuna-based search \cite{63} on tiles drawn from STARC-9 \cite{62} and the curated artifact dataset to balance tissue preservation and artifact suppression. }}
	\label{fig:8}       
\end{figure*} 

\subsection{Phase 1: Pen-mark removal}
Pen marks \cite{33} are non-biological ink annotations applied to glass slides during routine pathology workflows to indicate ROI or areas to avoid during analysis. Because these annotations are present before slide digitization, they appear in the scanned WSI as high-contrast structures that can enter downstream tile extraction pipelines. In routine collections, pen ink appeared in multiple colors \cite{19} (most commonly, red, blue, green, and black), with substantial variation across institutions, pen types and stroke thickness. Such regions are particularly problematic in weakly supervised MIL settings because artifact-dominant tiles can dilute bag representations in pooling-based models and attract high attention, even when the slide-level prediction is correct (Fig.~\ref{fig:5}). To prevent pen-mark dominant regions from entering downstream analysis, Phase 1 performs slide-level pen masking before tiling. Candidate ink pixels are identified using conservative RGB color thresholding \cite{61} to generate a binary pen mask which is excluded from subsequent processing. Fixed-size tiles ($256 \times 256$ pixels at magnification level 0) are then extracted only from the remaining unmasked slide area at the highest resolution level used for analysis. Tiles rejected at this stage are assigned to the Set4 category.  

\subsection{Phase 2: Optical-density analysis and quality filtering}
After pen mark removal, Phase 2 operates at the tile level to remove tiles lacking reliable H\&E signal, including background-dominant regions, dust or debris, and weakly stained areas. RGB intensities were transformed into the optical-density space using logarithmic mapping, enabling stain-aware analysis that is less sensitive to scanner illumination and color variation. Tiles were retained only when the mean eosin-channel intensity exceeded 0.01, a conservative threshold selected to suppress background-dominant tiles while preserving lightly stained tissue across heterogeneous cohorts. To further exclude artifacts and non-tissue background regions in an architecture-agnostic manner, several lightweight quality filters were applied (Fig.~\ref{fig:8}A). Residual preparation artifacts were suppressed by excluding tiles in which dark regions, such as black histology ink or dense tissue folds, occupied more than 25\% of the tile area. Blur was assessed using Sobel-derived edge content, under the assumption that blurred tiles lack sufficient edge structure; tiles with an edge score below 3\% were therefore discarded. Uniform regions (similar pixel range regions) were identified by applying a 50×50-pixel window (non-overlapping) across each tile and evaluating local intensity variance; tiles were removed when such uniform regions exceeded 70\% of the tile area. Finally, tiles with a background proportion greater than 85\% were discarded, where background was defined using a predefined white-pixel intensity range. Tiles that clearly failed these criteria were assigned to the Set4 category, while tiles passing all filters advanced to Phase 3. These thresholds were selected using a data-driven procedure on two complementary data sources. First, the STARC-9 \cite{62} training and validation sets were used, spanning both TCGA and Stanford WSI, multiple organs, scanners, and staining conditions. Although STARC-9 is a CRC dataset, the thresholds used in this phase are derived from pixel-level statistics (mean eosin, dark region, edge score, and uniform region) rather than organ-specific morphology. These filters operate on properties governed by H\&E staining physics, which are largely consistent across tissue types. Because the criteria assess stain distribution and spatial intensity variation rather than histologic patterns, the resulting thresholds generalize across organs and datasets. Second, an artifact-focused tile collection of 15,000 examples was curated to cover representative uninformative patterns, including pen marks, dust or debris, air bubbles, blur, tissue folds, scanner artifacts, and stain artifacts. Candidate thresholds were optimized using Optuna \cite{63} over predefined ranges, jointly balancing artifact suppression and tissue retention. Configurations were evaluated by minimizing artifact carryover while preserving diagnostically relevant tissue, using tile-retention statistics and the stability of downstream WSI-level performance. The final thresholds corresponded to Pareto-efficient operating points that generalized across cohorts rather than to dataset-specific optima. 

\subsection{Phase 3: Autoencoder-based utility filtering}
Although Phases 1 and 2 remove a substantial fraction of obvious artifacts, visually subtle artifacts, including dust/debris, mild blur, weak pen mark remnants, low-contrast non-tissue regions, and tiles with limited tissue content can still persist. These regions often do not violate deterministic threshold criteria strongly enough to be excluded. Phase 3 therefore introduces a representation-driven filtering step to suppress such artifact tiles while preserving diagnostically meaningful tissue. Phase 3 uses an unsupervised convolutional autoencoder trained explicitly on uninformative tiles. Unlike conventional reconstruction-based approaches that model normal tissue appearance, this design was intended to reconstruct artifact patterns with low reconstruction loss, such that biologically meaningful tissue yields comparatively higher reconstruction loss. Training data were curated from 40 WSI spanning Stanford (20), TCGA-LUAD (5), and CAMELYON16 (15), yielding approximately 15,000 tiles using DeepCluster++ \cite{62} containing representative uninformative tiles, including non-tissue background, air bubbles, dust or debris, pen marks, tissue folds, blur, and low tissue-content regions. High-quality tissue tiles were intentionally excluded from training to bias the autoencoder toward reconstruction of artifact-like patterns. The autoencoder operated on 256×256 RGB tiles and consisted of a lightweight convolutional neural network (CNN) encoder-decoder with three downsampling layers and symmetric upsampling blocks (Fig.~\ref{fig:8}B). Training proceeded until reconstruction stabilized, monitored using structural similarity index (SSIM) and peak signal-to-noise ratio (PSNR). 

During inference, each tile was assigned a reconstruction loss ($L$) together with its uniform-region score proportion estimated in Phase 2, and tiles were stratified into four categories: Set1, Set2, Set3, and Set4. High-utility tiles ($L \ge t_1$ and uniform-region score $< 30\%$) were assigned to Set1 and comprised tissue-dense regions considered relevant for downstream analysis. Tiles satisfying $t_2 \le L \le t_1$ with a uniform-region score of $30\text{--}50\%$ were assigned to Set2, whereas tiles satisfying $t_3 \le L \le t_2$ with a uniform-region score $> 50\%$ were assigned to Set3. 
Because Set2 contained partially tissue-filled regions and Set3 predominantly captured low-structural-density areas, including adipose-rich or background-dominant tissue compartments, both sets were categorized as low-utility tiles. Tiles rejected by MUFASA ($L < t_3$) were assigned to Set4 (uninformative), which comprised non-tissue background, artifacts, and regions with minimal tissue content. These tiles were considered uninformative for the primary WSI-level tasks because they contributed little task-relevant signal. Representative tiles from these categories are shown in Fig.~\ref{fig:2}B. Because the autoencoder was optimized to reconstruct artifact-like patterns, tiles containing tumor, stroma, epithelium, or lymphocyte-rich regions typically yielded higher reconstruction error and were preferentially assigned to Sets 1–3 (Fig.~\ref{fig:8}C–E), whereas artifact tiles that escaped Phases 1 and 2 were reconstructed with lower reconstruction loss and assigned to Set4 (Fig.~\ref{fig:8}F). In the present study, Set1 was used for the primary WSI-level analyses because it contained the highest signal-to-noise subset of tissue-dense tiles. Set2 and Set3 were retained as optional outputs for boundary-sensitive or tile-level applications, including tissue-composition analysis and dataset curation, but were excluded from the principal WSI-level classification and survival experiments to avoid introducing low-utility instances into MIL aggregation.

The loss thresholds ($t_1$=0.002,$t_2$=0.0015, $t_3$=0.001) were selected using a data-driven ablation procedure rather than manual tuning (Fig.~\ref{fig:8}H). Candidate threshold ranges were optimized using Optuna \cite{63} over the STARC-9 dataset \cite{62} and the curated artifact-focused training set, with approximately 50,000 tiles per set used to characterize stable operating regions. Threshold selection was guided by Pareto-efficient trade-offs between artifact suppression and tissue preservation. In practice, increasing or decreasing each threshold shifted tiles between adjacent categories in predictable ways: raising $t_1$ reduced assignment to Set1 and increased spillover into Set2; varying $t_2$ controlled the boundary between Set2 and Set3 categories; and varying $t_3$ determined whether Set3 tiles were retained or collapsed into Set4. The selected thresholds corresponded to operating points that maximized correct placement of adjacent categories simultaneously, rather than maximizing performance on any single dataset or downstream task.  

\subsection{Phase 4: Content-aware tissue recovery via lightweight classifier}
Although Phase 2 removed a large fraction of uninformative tiles, certain biologically relevant tissue types remained vulnerable to over-filtering, particularly adipose and mucin-rich regions. These tissues are characterized by low optical density, weak stain contrast, and large optically clear areas, which make them visually similar to background, dust or debris, or scanner artifacts. As a result, some of these tiles could be misclassified as artifact during optical-density screening. To reduce these systematic false negatives, Phase 4 introduced an optional lightweight content-aware tissue classifier (Fig.~\ref{fig:8}G). The classifier performs four-way classification on $256 \times 256$ RGB tiles: mucin, adipose, other tissue, and artifact. Adipose and mucin examples were obtained from the STARC-9 dataset \cite{62}, whereas “other tissue” and artifact examples were curated from tiles rejected during Phase 2 across diverse Stanford and TCGA cohorts. To promote morphologic diversity while avoiding redundant sampling, a fixed budget of 5,000 tiles per class was selected using DeepCluster++ \cite{62}. Tile features were extracted using a ResNet-50 \cite{45} at the third residual block, yielding 1024-dimensional embeddings. During inference, only tiles rejected in Phase 2 were passed to this classifier. Tiles predicted as mucin or other tissue were restored in Set1, whereas tiles predicted as adipose were assigned to Set3. Tiles classified as artifact were moved to Set4. This “filter, then recover” design was chosen instead of a single universal supervised classifier, which would require exhaustive annotation across organ types, scanners, institutions, and artifact classes. The recovery classifier was restricted to a narrower subset of borderline tiles and targeted only the tissue classes most susceptible to false-negative filtering. This reduced the annotation burden while preserving extensibility; if new recoverable tissue patterns become clinically relevant, additional classes can be incorporated without redesigning the earlier phases. 

After refinement, all retained tiles were remapped to their original spatial coordinates to generate category-specific masks for Sets1–3, along with a corresponding aggregate mask. Representative WSI-level outputs using MUFASA on the TCGA-COAD, TCGA-BRCA, and the Stanford CRC cohorts are shown in Supplementary Fig.~\ref{fig:S2}, illustrating that the framework preserved the majority of tissue-containing regions while selectively suppressing uninformative regions. This representation allowed downstream pipelines to select task-specific subsets. 

\paragraph{\textbf{Data availability.}} The artifact dataset used to train the autoencoder will be made publicly available. The STARC-9 dataset used for threshold evaluation and training of the recovery classifier is already publicly available at: \url{https://huggingface.co/datasets/Path2AI/STARC-9/tree/main}. The TCGA data used in this study are publicly available at: \url{https://www.cancer.gov/ccg/research/genome-sequencing/tcga}, and the corresponding MUFASA-generated tissue masks will be made publicly available upon publication of the manuscript. The raw source data for generating the tables and figures are provided in the Supplementary materials.

\paragraph{\textbf{Code availability.}} The custom source code used in this study is currently being compiled for public release. The codebase will be made available to editors and reviewers upon request during the peer-review process and will be fully deposited in a public repository prior to final publication.

\paragraph{\textbf{Author contributions.}} R.J., J.S., and B.S. conducted study design and conceptualization of the framework. R.J. and B.S. performed software development, experimental setup, data analysis, and data visualization. J.S. provided critical input on experimental design, performance evaluation, and data analysis. J.S., R.J., B.S, S.N., M.N.P., and T.G. performed data acquisition and annotation. R.J. and J.S. wrote the manuscript. G.A.F., N.H.S., C.P.L., and T.J.M. reviewed and provided feedback on the manuscript. J.S. acquired funding and resources and supervised the project. G.A.F. provided additional resources. All authors reviewed and approved the final manuscript.

\paragraph{\textbf{Acknowledgments.}} We acknowledge financial support from the United States National Cancer Institute (NCI), National Institutes of Health (NIH) under award number R01 CA270437 to J.S. The content is solely the responsibility of the authors and does not necessarily represent the official views of the National Institutes of Health. S.N.’s contributions to this research were performed entirely while affiliated with and physically located at the Department of Pathology, Stanford University School of Medicine, as a visiting scholar. S.N.’s visiting scholarship was supported independently by a grant from the Korea Health Industry Development Institute (KHIDI), Ministry of Health \& Welfare, Republic of Korea, which had no role in the design, execution, or funding of this project.

\paragraph{\textbf{Competing interests.}} The authors declare no competing interests.

\newpage 
\onecolumn
\begin{landscape}
\vspace*{-47em}
\begin{center} 
    {\LARGE \textbf{Supplementary Materials}}  \\[1em] 
\end{center}

\begin{table}[H]
\centering
\caption{\footnotesize \textbf{Tumor-versus-no-tumor classification on CAMELYON16 under four preprocessing pipelines.} Performance is reported as mean$\pm$s.d. across five random runs for accuracy (ACC), balanced accuracy (BACC), Macro-AUC, and Macro-F1. All MIL models used identical feature extraction and training settings; only the preprocessing pipeline used to generate tissue masks differed (CLAM, Histolab, Trident, or MUFASA). MUFASA achieved the strongest overall average performance and the most consistent gains in Macro-AUC across architectures, indicating improved class-level discrimination under artifact-aware preprocessing. The best-performing preprocessor for each metric, averaged across models, is highlighted in bold, and the second-best is underlined (bottom row).}
\label{tab:1}
\resizebox{1.4\textwidth}{!}{%
\begin{tabular}{l|cccc|cccc|cccc|cccc}
\toprule
\multirow{2}{*}{\textbf{Model}} 
    & \multicolumn{4}{c|}{\textbf{CLAM}} 
    & \multicolumn{4}{c|}{\textbf{Histolab}} 
    & \multicolumn{4}{c|}{\textbf{Trident}} 
    & \multicolumn{4}{c}{\textbf{MUFASA}} \\
\cmidrule(lr){2-5}\cmidrule(lr){6-9}\cmidrule(lr){10-13}\cmidrule(lr){14-17}
    & ACC & BACC & Macro-AUC & Macro-F1 
    & ACC & BACC & Macro-AUC & Macro-F1 
    & ACC & BACC & Macro-AUC & Macro-F1 
    & ACC & BACC & Macro-AUC & Macro-F1 \\
\midrule
Mean-pooling      & 60.53$\pm$8.26  & 55.16$\pm$2.72  & 59.73$\pm$2.09  & 50.09$\pm$4.83   & 56.43$\pm$8.11  & 53.01$\pm$2.43  & 53.62$\pm$0.70  & 49.71$\pm$6.28   & 61.48$\pm$1.64  & 53.14$\pm$1.18  & 53.11$\pm$0.94  & 48.41$\pm$3.08   & 65.89$\pm$2.53  & 56.83$\pm$3.20  & 62.90$\pm$3.11  & 53.60$\pm$6.03  \\
DeepAttnMISL & 57.20$\pm$9.61  & 50.00$\pm$0.00  & 67.75$\pm$1.14  & 36.12$\pm$4.30   & 57.20$\pm$9.61  & 50.00$\pm$0.00  & 53.49$\pm$4.49  & 36.12$\pm$4.30   & 52.40$\pm$11.77 & 50.00$\pm$0.00  & 53.97$\pm$3.14  & 33.97$\pm$5.27   & 62.01$\pm$0.00  & 50.00$\pm$0.00  & 69.54$\pm$1.01  & 38.27$\pm$0.00  \\
CLAM-MB           & 82.01$\pm$1.24  & 78.54$\pm$0.82  & 84.11$\pm$1.24  & 79.77$\pm$1.10   & 78.75$\pm$1.44  & 75.75$\pm$0.95  & 79.60$\pm$0.48  & 76.55$\pm$1.22   & 79.84$\pm$1.77  & 76.15$\pm$1.09  & 80.34$\pm$0.81  & 77.38$\pm$1.61   & 84.03$\pm$0.93  & 80.95$\pm$1.02  & 87.38$\pm$1.42  & 82.15$\pm$0.93  \\
DSMIL             & 69.74$\pm$6.48  & 61.41$\pm$9.55  & 70.30$\pm$1.07  & 56.72$\pm$15.27  & 68.50$\pm$5.57  & 60.62$\pm$8.88  & 63.30$\pm$6.89  & 56.23$\pm$14.84  & 68.68$\pm$5.49  & 59.80$\pm$8.11  & 65.74$\pm$7.98  & 55.18$\pm$13.91  & 70.23$\pm$6.87  & 62.22$\pm$9.84  & 73.87$\pm$4.99  & 58.75$\pm$14.99 \\
TransMIL          & 81.05$\pm$2.36  & 80.62$\pm$0.59  & 88.68$\pm$2.99  & 80.59$\pm$2.29   & 76.03$\pm$6.79  & 75.63$\pm$4.80  & 84.69$\pm$3.59  & 74.92$\pm$6.62   & 78.06$\pm$3.23  & 78.10$\pm$2.42  & 87.30$\pm$1.77  & 77.16$\pm$2.78   & 83.25$\pm$2.28  & 82.38$\pm$1.81  & 90.63$\pm$0.56  & 82.63$\pm$1.82  \\
DTFD-MIL          & 81.70$\pm$2.06  & 78.52$\pm$1.46  & 83.14$\pm$0.30  & 79.60$\pm$1.87   & 78.91$\pm$0.76  & 75.48$\pm$0.61  & 78.44$\pm$0.17  & 76.44$\pm$0.72   & 77.67$\pm$0.90  & 73.93$\pm$1.03  & 78.46$\pm$0.72  & 74.87$\pm$1.07   & 84.80$\pm$0.79  & 81.81$\pm$1.60  & 87.60$\pm$2.45  & 83.01$\pm$1.17  \\
MHIM-MIL          & 85.11$\pm$2.47  & 81.83$\pm$4.09  & 91.79$\pm$0.84  & 83.00$\pm$3.73   & 84.97$\pm$2.31  & 81.87$\pm$3.53  & 90.76$\pm$1.18  & 82.80$\pm$2.91   & 83.81$\pm$1.92  & 80.43$\pm$4.20  & 90.08$\pm$2.86  & 81.24$\pm$3.61   & 87.90$\pm$1.87  & 85.26$\pm$3.07  & 92.96$\pm$0.65  & 86.46$\pm$2.60  \\
Additive MIL      & 79.41$\pm$1.19  & 76.30$\pm$1.10  & 81.38$\pm$0.85  & 77.57$\pm$1.76   & 77.86$\pm$0.79  & 75.95$\pm$1.63  & 80.15$\pm$1.95  & 75.10$\pm$0.96   & 78.59$\pm$1.08  & 76.40$\pm$1.18  & 80.53$\pm$1.77  & 76.01$\pm$0.77   & 80.77$\pm$2.25  & 77.89$\pm$1.51  & 81.94$\pm$0.91  & 79.21$\pm$2.10  \\
DGR-MIL           & 79.38$\pm$9.38  & 74.99$\pm$13.21 & 83.09$\pm$9.82  & 73.26$\pm$17.94  & 74.72$\pm$9.22  & 71.95$\pm$11.84 & 81.58$\pm$8.65  & 69.18$\pm$16.60  & 66.97$\pm$18.15 & 63.80$\pm$16.91 & 77.20$\pm$11.85 & 54.91$\pm$25.07  & 81.72$\pm$3.29  & 77.90$\pm$6.96  & 89.48$\pm$1.28  & 75.45$\pm$10.00 \\
ILRA-MIL          & 62.01$\pm$0.00  & 50.00$\pm$0.00  & 68.78$\pm$2.17  & 38.27$\pm$0.00   & 61.52$\pm$0.41  & 46.96$\pm$3.51  & 59.71$\pm$1.20  & 37.77$\pm$0.68   & 61.74$\pm$0.34  & 48.63$\pm$1.98  & 61.19$\pm$0.34  & 36.39$\pm$2.16   & 62.01$\pm$0.00  & 50.00$\pm$0.00  & 69.68$\pm$1.49  & 38.27$\pm$0.00  \\
ACMIL             & 80.61$\pm$1.47  & 78.68$\pm$1.49  & 83.86$\pm$1.82  & 79.09$\pm$1.53   & 80.15$\pm$2.80  & 78.14$\pm$2.10  & 81.89$\pm$1.02  & 78.28$\pm$2.58   & 79.37$\pm$0.79  & 76.33$\pm$0.77  & 79.62$\pm$0.81  & 77.42$\pm$0.82   & 81.94$\pm$1.80  & 80.62$\pm$2.79  & 86.52$\pm$1.79  & 80.04$\pm$1.49  \\
MambaMIL      & 82.95$\pm$4.47  & 78.46$\pm$5.73  & 86.56$\pm$2.41  & 80.22$\pm$6.49   & 83.16$\pm$3.12  & 79.86$\pm$3.35  & 85.19$\pm$1.83  & 81.15$\pm$3.55   & 84.69$\pm$3.11  & 81.01$\pm$2.71  & 86.18$\pm$2.53  & 82.66$\pm$3.02   & 86.60$\pm$1.40  & 83.65$\pm$1.33  & 88.86$\pm$1.27  & 85.55$\pm$1.37  \\
TDA-MIL           & 84.03$\pm$2.67  & 81.11$\pm$4.22  & 88.36$\pm$1.02  & 81.94$\pm$3.91   & 84.34$\pm$1.73  & 82.07$\pm$2.46  & 89.26$\pm$1.09  & 82.74$\pm$2.14   & 82.94$\pm$2.30  & 79.92$\pm$2.07  & 88.10$\pm$1.71  & 80.85$\pm$2.55   & 85.01$\pm$1.86  & 83.63$\pm$1.31  & 91.58$\pm$1.29  & 84.52$\pm$1.89  \\
\midrule
\textbf{Average}  & \underline{75.82} & \underline{71.20} & \underline{79.81} & \underline{68.94} & 74.04 & 69.79 & 75.51 & 67.49 & 73.55 & 69.08 & 75.52 & 65.88 & \textbf{78.16} & \textbf{73.31} & \textbf{82.53} & \textbf{71.35} \\
\bottomrule
\end{tabular}%
}
\end{table}

\begin{table}[H]
\centering
\caption{\footnotesize \textbf{Tumor subtyping of NSCLC (TCGA-LUAD versus TCGA-LUSC) under four preprocessing pipelines.} Performance is reported as mean$\pm$s.d. across five cross-validation runs for accuracy (ACC), balanced accuracy (BACC), Macro-AUC, and Macro-F1. All MIL models were trained using identical feature extraction and optimization settings; only the preprocessing pipeline differed. MUFASA produced the best overall performance across architectures, supporting improved subtype discrimination under artifact-aware tissue selection. The best-performing preprocessor for each metric, averaged across models, is highlighted in bold, and the second-best is underlined (bottom row).}
\label{tab:2}
\resizebox{1.4\textwidth}{!}{%
\begin{tabular}{l|cccc|cccc|cccc|cccc}
\toprule
\multirow{2}{*}{\textbf{Model}} 
    & \multicolumn{4}{c|}{\textbf{CLAM}} 
    & \multicolumn{4}{c|}{\textbf{Histolab}} 
    & \multicolumn{4}{c|}{\textbf{Trident}} 
    & \multicolumn{4}{c}{\textbf{MUFASA}} \\
\cmidrule(lr){2-5}\cmidrule(lr){6-9}\cmidrule(lr){10-13}\cmidrule(lr){14-17}
    & ACC & BACC & Macro-AUC & Macro-F1
    & ACC & BACC & Macro-AUC & Macro-F1
    & ACC & BACC & Macro-AUC & Macro-F1
    & ACC & BACC & Macro-AUC & Macro-F1 \\
\midrule
Mean-pooling      & 81.48$\pm$0.60  & 81.04$\pm$0.69  & 88.68$\pm$1.40  & 81.12$\pm$0.62   & 80.66$\pm$0.81  & 80.12$\pm$1.26  & 88.22$\pm$1.39  & 80.25$\pm$1.05   & 80.30$\pm$1.83  & 79.91$\pm$1.84  & 88.23$\pm$1.93  & 79.98$\pm$1.83   & 82.85$\pm$1.10  & 82.53$\pm$1.13  & 90.13$\pm$1.39  & 82.49$\pm$1.17  \\
DeepAttnMISL  & 70.93$\pm$3.69  & 69.41$\pm$3.67  & 76.97$\pm$3.10  & 69.52$\pm$3.79   & 71.05$\pm$3.42  & 69.55$\pm$3.39  & 76.64$\pm$3.30  & 69.66$\pm$3.48   & 71.28$\pm$4.16  & 69.83$\pm$4.25  & 76.98$\pm$2.62  & 69.93$\pm$4.42   & 75.27$\pm$3.54  & 73.97$\pm$3.50  & 81.17$\pm$2.89  & 74.24$\pm$3.58  \\
CLAM-MB           & 86.28$\pm$1.63  & 85.96$\pm$1.88  & 93.13$\pm$1.12  & 85.90$\pm$1.39   & 86.20$\pm$1.69  & 85.66$\pm$1.88  & 93.20$\pm$1.07  & 86.17$\pm$1.28   & 84.96$\pm$1.86  & 85.47$\pm$1.84  & 93.12$\pm$1.15  & 85.44$\pm$1.70   & 88.69$\pm$1.80  & 88.08$\pm$1.91  & 94.92$\pm$1.16  & 88.12$\pm$1.93  \\
DSMIL             & 80.66$\pm$3.70  & 79.76$\pm$3.99  & 86.22$\pm$2.92  & 79.98$\pm$4.09   & 80.43$\pm$3.97  & 79.43$\pm$4.33  & 86.58$\pm$3.04  & 79.69$\pm$4.40   & 78.44$\pm$3.86  & 77.37$\pm$4.58  & 83.86$\pm$4.92  & 77.43$\pm$4.62   & 83.72$\pm$3.52  & 82.52$\pm$3.18  & 89.12$\pm$0.64  & 83.13$\pm$2.99  \\
TransMIL          & 87.27$\pm$1.56  & 87.15$\pm$2.03  & 94.20$\pm$0.26  & 87.59$\pm$1.85   & 82.93$\pm$4.24  & 81.64$\pm$5.29  & 93.81$\pm$0.52  & 81.52$\pm$5.23   & 86.34$\pm$2.48  & 86.09$\pm$3.33  & 93.90$\pm$0.54  & 86.94$\pm$3.36   & 89.40$\pm$0.83  & 89.63$\pm$1.38  & 95.96$\pm$0.52  & 89.89$\pm$0.87  \\
DTFD-MIL          & 86.64$\pm$1.52  & 86.40$\pm$1.44  & 94.00$\pm$0.93  & 86.05$\pm$1.47   & 86.74$\pm$2.01  & 86.66$\pm$1.89  & 93.76$\pm$1.21  & 86.75$\pm$1.96   & 86.27$\pm$1.83  & 86.27$\pm$1.80  & 93.69$\pm$1.39  & 86.33$\pm$1.88   & 88.84$\pm$1.85  & 88.20$\pm$2.07  & 95.14$\pm$0.83  & 88.64$\pm$1.89  \\
MHIM-MIL          & 87.72$\pm$2.16  & 86.59$\pm$2.75  & 95.32$\pm$0.88  & 86.87$\pm$2.07   & 85.66$\pm$3.35  & 85.09$\pm$3.33  & 94.91$\pm$0.99  & 85.65$\pm$1.99   & 86.72$\pm$1.97  & 86.30$\pm$2.62  & 94.64$\pm$0.84  & 86.06$\pm$2.57   & 89.98$\pm$0.97  & 89.72$\pm$1.04  & 96.60$\pm$0.63  & 89.71$\pm$0.60  \\
Additive MIL      & 86.16$\pm$3.02  & 85.29$\pm$2.87  & 93.39$\pm$1.12  & 85.90$\pm$3.00   & 86.76$\pm$2.28  & 86.59$\pm$2.47  & 93.57$\pm$1.25  & 86.62$\pm$3.33   & 86.42$\pm$4.65  & 85.90$\pm$2.27  & 92.87$\pm$2.12  & 86.42$\pm$2.49   & 88.37$\pm$2.43  & 87.69$\pm$2.36  & 94.35$\pm$1.27  & 88.15$\pm$2.46  \\
DGR-MIL           & 75.17$\pm$14.06 & 74.44$\pm$14.83 & 91.35$\pm$2.33  & 70.29$\pm$20.62  & 74.74$\pm$15.31 & 74.57$\pm$12.79 & 87.63$\pm$4.53  & 71.67$\pm$20.64  & 75.90$\pm$15.73 & 76.08$\pm$13.21 & 89.10$\pm$4.31  & 72.46$\pm$20.81  & 89.06$\pm$0.88  & 88.87$\pm$1.21  & 95.12$\pm$1.18  & 88.68$\pm$1.25  \\
ILRA-MIL          & 87.84$\pm$2.49  & 87.52$\pm$2.65  & 94.50$\pm$1.18  & 87.04$\pm$2.04   & 87.18$\pm$2.96  & 88.01$\pm$2.12  & 93.91$\pm$1.27  & 86.74$\pm$2.34   & 87.01$\pm$2.60  & 86.72$\pm$2.51  & 93.99$\pm$1.28  & 86.14$\pm$2.11   & 89.91$\pm$1.55  & 89.67$\pm$1.26  & 95.56$\pm$1.07  & 89.37$\pm$1.20  \\
ACMIL             & 87.93$\pm$2.33  & 87.49$\pm$2.38  & 94.20$\pm$1.49  & 87.71$\pm$2.45   & 87.58$\pm$2.33  & 87.24$\pm$2.87  & 94.27$\pm$1.09  & 87.29$\pm$2.55   & 84.01$\pm$3.59  & 83.70$\pm$4.05  & 93.70$\pm$1.48  & 83.95$\pm$4.01   & 89.08$\pm$2.00  & 89.21$\pm$1.75  & 95.10$\pm$0.86  & 89.35$\pm$1.70  \\
MambaMIL      & 86.52$\pm$7.24  & 85.67$\pm$8.36  & 94.96$\pm$0.98  & 85.44$\pm$8.74   & 87.87$\pm$2.77  & 87.85$\pm$2.37  & 94.26$\pm$1.03  & 87.33$\pm$3.07   & 87.38$\pm$2.27  & 87.36$\pm$2.88  & 94.11$\pm$0.85  & 86.64$\pm$2.89   & 88.36$\pm$2.34  & 88.59$\pm$1.79  & 96.18$\pm$1.01  & 88.10$\pm$2.12  \\
TDA-MIL           & 86.98$\pm$0.61  & 87.51$\pm$1.38  & 95.22$\pm$0.53  & 87.60$\pm$1.78   & 85.90$\pm$2.23  & 86.14$\pm$3.10  & 94.48$\pm$0.33  & 85.53$\pm$2.52   & 85.82$\pm$1.52  & 86.21$\pm$2.03  & 94.48$\pm$0.78  & 85.88$\pm$1.55   & 89.39$\pm$0.74  & 89.33$\pm$0.69  & 96.38$\pm$0.52  & 89.34$\pm$0.95  \\
\midrule
\textbf{Average}  & \underline{83.96} & \underline{83.43} & \underline{91.70} & \underline{83.15}   & 83.36 & 82.96 & 91.17 & 82.67   & 83.15 & 82.86 & 90.97 & 82.58   & \textbf{87.18} & \textbf{86.79} & \textbf{93.54} & \textbf{86.86} \\
\bottomrule
\end{tabular}%
}
\end{table}

\begin{table}[t!]
\centering
\caption{\footnotesize \textbf{MSS/MSI prediction on the Stanford colorectal cancer cohort under four preprocessing pipelines.} Performance is reported as mean$\pm$s.d. across five cross-validation runs for accuracy (ACC), balanced accuracy (BACC), Macro-AUC, and Macro-F1. Downstream feature extraction and MIL training settings were held constant across all comparisons. MUFASA yielded the strongest and most consistent improvements across architectures, indicating improved molecular biomarker prediction with reduced artifact carryover. The best-performing preprocessor for each metric, averaged across models, is highlighted in bold, and the second-best is underlined (bottom row).}
\label{tab:3}
\resizebox{1.4\textwidth}{!}{%
\begin{tabular}{l|cccc|cccc|cccc|cccc}
\toprule
\multirow{2}{*}{\textbf{Model}} 
    & \multicolumn{4}{c|}{\textbf{CLAM}} 
    & \multicolumn{4}{c|}{\textbf{Histolab}} 
    & \multicolumn{4}{c|}{\textbf{Trident}} 
    & \multicolumn{4}{c}{\textbf{MUFASA}} \\
\cmidrule(lr){2-5}\cmidrule(lr){6-9}\cmidrule(lr){10-13}\cmidrule(lr){14-17}
    & ACC & BACC & Macro-AUC & Macro-F1
    & ACC & BACC & Macro-AUC & Macro-F1
    & ACC & BACC & Macro-AUC & Macro-F1
    & ACC & BACC & Macro-AUC & Macro-F1 \\
\midrule
Mean-pooling      & 86.97$\pm$0.62  & 59.15$\pm$1.29  & 82.00$\pm$2.16  & 61.97$\pm$1.81   & 86.07$\pm$0.69  & 57.22$\pm$2.04  & 80.88$\pm$2.39  & 58.76$\pm$3.11   & 85.86$\pm$0.78  & 55.88$\pm$2.55  & 80.01$\pm$2.88  & 56.68$\pm$4.26   & 87.33$\pm$0.79  & 60.72$\pm$1.72  & 83.31$\pm$1.91  & 63.86$\pm$2.45  \\
DeepAttnMISL  & 84.84$\pm$0.26  & 50.00$\pm$0.00  & 72.85$\pm$2.08  & 45.90$\pm$0.07   & 84.88$\pm$0.22  & 50.00$\pm$0.00  & 71.51$\pm$1.35  & 45.91$\pm$0.06   & 84.76$\pm$0.34  & 50.00$\pm$0.00  & 70.13$\pm$3.14  & 45.89$\pm$0.11   & 84.88$\pm$0.22  & 50.00$\pm$0.00  & 74.47$\pm$2.60  & 45.91$\pm$0.06  \\
CLAM-MB           & 85.57$\pm$1.29  & 61.36$\pm$3.73  & 82.53$\pm$2.92  & 63.81$\pm$4.41   & 85.54$\pm$1.47  & 58.67$\pm$1.62  & 80.21$\pm$2.42  & 60.78$\pm$2.21   & 85.24$\pm$1.18  & 58.27$\pm$1.84  & 78.19$\pm$3.48  & 60.14$\pm$2.42   & 87.93$\pm$0.86  & 67.59$\pm$4.28  & 84.15$\pm$3.97  & 70.69$\pm$3.59  \\
DSMIL             & 82.57$\pm$1.56  & 50.67$\pm$1.72  & 74.16$\pm$2.70  & 49.68$\pm$2.79   & 82.02$\pm$2.01  & 49.28$\pm$2.23  & 73.25$\pm$2.82  & 47.73$\pm$2.80   & 82.11$\pm$1.66  & 50.24$\pm$1.36  & 72.68$\pm$2.35  & 46.76$\pm$2.68   & 84.75$\pm$1.49  & 53.55$\pm$2.90  & 75.84$\pm$2.43  & 53.28$\pm$4.85  \\
TransMIL          & 82.33$\pm$4.38  & 64.14$\pm$8.32  & 79.52$\pm$0.66  & 62.31$\pm$8.82   & 79.34$\pm$8.58  & 66.04$\pm$6.37  & 78.47$\pm$1.31  & 64.13$\pm$5.60   & 86.21$\pm$2.48  & 63.14$\pm$6.86  & 76.81$\pm$0.70  & 61.71$\pm$6.94   & 85.55$\pm$1.46  & 70.52$\pm$3.40  & 82.21$\pm$1.45  & 69.63$\pm$2.76  \\
DTFD-MIL          & 84.77$\pm$1.35  & 56.17$\pm$1.93  & 82.75$\pm$2.15  & 57.83$\pm$3.35   & 84.35$\pm$1.27  & 55.29$\pm$1.92  & 80.88$\pm$1.44  & 56.62$\pm$3.09   & 84.63$\pm$1.47  & 55.48$\pm$0.79  & 80.75$\pm$2.23  & 56.42$\pm$1.19   & 86.41$\pm$1.09  & 60.91$\pm$2.73  & 83.98$\pm$2.15  & 63.74$\pm$3.27  \\
MHIM-MIL          & 89.49$\pm$1.79  & 77.73$\pm$3.96  & 90.26$\pm$3.24  & 79.61$\pm$2.27   & 88.26$\pm$2.40  & 73.45$\pm$4.54  & 89.76$\pm$3.18  & 74.31$\pm$3.82   & 88.37$\pm$0.99  & 72.27$\pm$3.26  & 89.58$\pm$2.69  & 74.48$\pm$1.83   & 90.06$\pm$1.83  & 79.07$\pm$4.38  & 91.41$\pm$2.78  & 80.26$\pm$2.63  \\
Additive MIL      & 86.91$\pm$1.94  & 67.79$\pm$5.46  & 83.07$\pm$3.72  & 71.02$\pm$5.81   & 85.61$\pm$0.99  & 65.95$\pm$5.48  & 80.91$\pm$4.33  & 68.62$\pm$5.40   & 85.93$\pm$1.30  & 63.13$\pm$2.98  & 80.70$\pm$3.06  & 66.00$\pm$3.28   & 87.92$\pm$1.49  & 69.81$\pm$3.89  & 85.73$\pm$2.27  & 73.56$\pm$3.29  \\
DGR-MIL           & 80.73$\pm$3.20  & 64.89$\pm$8.92  & 82.03$\pm$2.68  & 63.30$\pm$9.59   & 40.27$\pm$22.12 & 61.02$\pm$7.54  & 81.74$\pm$3.28  & 36.12$\pm$17.25  & 66.69$\pm$26.53 & 50.00$\pm$0.00  & 78.67$\pm$4.26  & 39.21$\pm$13.35  & 83.88$\pm$2.42  & 71.75$\pm$4.25  & 85.65$\pm$2.59  & 71.15$\pm$4.19  \\
ILRA-MIL          & 86.36$\pm$1.26  & 66.73$\pm$3.03  & 85.33$\pm$2.57  & 69.27$\pm$3.17   & 85.61$\pm$1.17  & 63.51$\pm$2.94  & 84.86$\pm$2.50  & 65.92$\pm$3.07   & 84.84$\pm$1.06  & 61.44$\pm$1.38  & 82.60$\pm$2.91  & 63.48$\pm$1.33   & 87.66$\pm$0.82  & 71.47$\pm$4.12  & 86.51$\pm$3.12  & 73.22$\pm$2.98  \\
ACMIL             & 87.50$\pm$0.86  & 73.19$\pm$2.49  & 87.82$\pm$2.68  & 74.30$\pm$0.49   & 86.87$\pm$1.19  & 71.12$\pm$1.01  & 86.45$\pm$3.15  & 73.39$\pm$1.86   & 87.69$\pm$0.66  & 70.32$\pm$3.09  & 85.89$\pm$3.39  & 73.60$\pm$3.48   & 88.73$\pm$0.65  & 78.94$\pm$1.16  & 88.90$\pm$2.59  & 77.09$\pm$1.40  \\
MambaMIL      & 88.36$\pm$1.33  & 72.27$\pm$3.72  & 91.82$\pm$2.18  & 75.25$\pm$3.35   & 88.19$\pm$1.12  & 72.64$\pm$3.24  & 91.45$\pm$2.25  & 73.54$\pm$4.91   & 87.30$\pm$0.81  & 69.43$\pm$5.03  & 90.75$\pm$2.90  & 71.64$\pm$3.27   & 90.85$\pm$0.77  & 80.79$\pm$2.92  & 93.33$\pm$2.21  & 81.32$\pm$1.74  \\
TDA-MIL           & 88.63$\pm$1.04  & 72.64$\pm$4.22  & 90.57$\pm$2.42  & 75.47$\pm$4.03   & 87.05$\pm$2.66  & 72.00$\pm$3.63  & 89.49$\pm$2.14  & 76.44$\pm$2.50   & 87.72$\pm$2.49  & 71.41$\pm$3.25  & 89.74$\pm$1.95  & 74.03$\pm$2.79   & 90.06$\pm$1.83  & 78.14$\pm$1.44  & 92.23$\pm$1.78  & 78.68$\pm$1.43  \\
\midrule
\textbf{Average}  & \underline{85.77} & \underline{64.36} & \underline{83.43} & \underline{65.36}   & 81.85 & 62.78 & 82.29 & 61.71   & 84.15 & 60.84 & 81.26 & 60.77   & \textbf{87.38} & \textbf{68.68} & \textbf{85.14} & \textbf{69.41} \\
\bottomrule
\end{tabular}%
}
\end{table}

\begin{table}[t!]
\centering
\caption{\footnotesize \textbf{MSS/MSI prediction on the TCGA colorectal cancer cohort under four preprocessing pipelines.} Performance is reported as mean$\pm$s.d. across five cross-validation runs for accuracy (ACC), balanced accuracy (BACC), Macro-AUC, and Macro-F1. All MIL models used identical feature extraction and training configurations, with preprocessing as the only varying component. MUFASA consistently improved performance relative to baseline preprocessing pipelines, supporting its generalizability across independent colorectal cancer cohorts. The best-performing preprocessor for each metric, averaged across models, is highlighted in bold, and the second-best is underlined (bottom row).}
\label{tab:4}
\resizebox{1.4\textwidth}{!}{%
\begin{tabular}{l|cccc|cccc|cccc|cccc}
\toprule
\multirow{2}{*}{\textbf{Model}} 
    & \multicolumn{4}{c|}{\textbf{CLAM}} 
    & \multicolumn{4}{c|}{\textbf{Histolab}} 
    & \multicolumn{4}{c|}{\textbf{Trident}} 
    & \multicolumn{4}{c}{\textbf{MUFASA}} \\
\cmidrule(lr){2-5}\cmidrule(lr){6-9}\cmidrule(lr){10-13}\cmidrule(lr){14-17}
    & ACC & BACC & Macro-AUC & Macro-F1
    & ACC & BACC & Macro-AUC & Macro-F1
    & ACC & BACC & Macro-AUC & Macro-F1
    & ACC & BACC & Macro-AUC & Macro-F1 \\
\midrule
Mean-pooling      & 82.51$\pm$0.80  & 50.91$\pm$1.82  & 64.93$\pm$3.81  & 46.86$\pm$3.47   & 82.51$\pm$0.80  & 50.91$\pm$1.82  & 66.19$\pm$3.38  & 46.86$\pm$3.47   & 81.85$\pm$0.96  & 50.51$\pm$1.57  & 65.44$\pm$4.11  & 46.53$\pm$3.07   & 82.58$\pm$0.76  & 50.91$\pm$1.82  & 69.07$\pm$3.03  & 46.96$\pm$3.42  \\
DeepAttnMISL  & 82.18$\pm$0.59  & 50.00$\pm$0.00  & 56.70$\pm$7.64  & 45.11$\pm$0.17   & 82.18$\pm$0.59  & 50.00$\pm$0.00  & 57.98$\pm$7.68  & 45.11$\pm$0.17   & 82.18$\pm$0.59  & 50.00$\pm$0.00  & 59.48$\pm$8.09  & 45.11$\pm$0.17   & 82.18$\pm$0.59  & 50.00$\pm$0.00  & 61.65$\pm$7.83  & 45.11$\pm$0.17  \\
CLAM-MB           & 82.10$\pm$1.21  & 51.61$\pm$3.76  & 69.47$\pm$7.15  & 48.25$\pm$6.51   & 82.11$\pm$1.19  & 51.14$\pm$2.03  & 69.37$\pm$7.22  & 49.10$\pm$6.72   & 82.84$\pm$1.94  & 53.33$\pm$3.43  & 70.57$\pm$5.43  & 51.56$\pm$6.19   & 83.24$\pm$1.18  & 54.06$\pm$3.68  & 75.81$\pm$5.36  & 53.68$\pm$6.66  \\
DSMIL             & 82.18$\pm$0.59  & 50.00$\pm$0.00  & 62.38$\pm$5.15  & 45.11$\pm$0.17   & 82.18$\pm$0.59  & 50.00$\pm$0.00  & 62.79$\pm$4.69  & 45.11$\pm$0.17   & 82.18$\pm$0.59  & 50.00$\pm$0.00  & 65.69$\pm$6.23  & 45.11$\pm$0.17   & 82.18$\pm$0.59  & 50.00$\pm$0.00  & 69.37$\pm$7.43  & 45.11$\pm$0.17  \\
TransMIL          & 79.25$\pm$4.02  & 57.35$\pm$6.25  & 71.98$\pm$4.72  & 56.37$\pm$12.30  & 81.45$\pm$1.28  & 52.65$\pm$1.81  & 70.71$\pm$2.60  & 53.50$\pm$9.56   & 80.21$\pm$2.64  & 53.06$\pm$5.84  & 61.11$\pm$4.69  & 49.76$\pm$7.18   & 84.18$\pm$1.23  & 59.35$\pm$5.36  & 72.29$\pm$3.81  & 59.10$\pm$11.26 \\
DTFD-MIL          & 82.18$\pm$0.59  & 50.00$\pm$0.00  & 64.17$\pm$4.68  & 45.11$\pm$0.17   & 81.51$\pm$1.56  & 50.00$\pm$0.00  & 65.73$\pm$4.90  & 44.86$\pm$0.56   & 81.85$\pm$0.96  & 49.80$\pm$0.40  & 66.01$\pm$3.69  & 45.01$\pm$0.29   & 82.53$\pm$0.82  & 50.91$\pm$1.82  & 69.53$\pm$5.01  & 45.11$\pm$0.17  \\
MHIM-MIL          & 79.21$\pm$2.41  & 58.30$\pm$5.02  & 80.23$\pm$6.99  & 57.17$\pm$6.39   & 76.88$\pm$3.75  & 55.47$\pm$4.96  & 79.38$\pm$3.44  & 54.76$\pm$6.52   & 79.68$\pm$3.24  & 62.96$\pm$6.07  & 82.14$\pm$5.23  & 59.75$\pm$4.06   & 82.54$\pm$0.78  & 65.33$\pm$8.06  & 86.10$\pm$2.69  & 63.35$\pm$2.91  \\
Additive MIL      & 79.80$\pm$2.58  & 52.95$\pm$3.76  & 70.33$\pm$5.98  & 52.12$\pm$7.73   & 78.82$\pm$4.23  & 51.85$\pm$2.88  & 70.40$\pm$6.88  & 52.01$\pm$7.42   & 79.58$\pm$2.70  & 52.22$\pm$3.66  & 71.51$\pm$7.04  & 50.58$\pm$7.31   & 81.14$\pm$1.23  & 53.13$\pm$3.48  & 74.84$\pm$8.84  & 53.02$\pm$8.00  \\
DGR-MIL           & 82.18$\pm$0.59  & 50.00$\pm$0.00  & 67.32$\pm$6.12  & 45.11$\pm$0.17   & 82.18$\pm$0.59  & 50.00$\pm$0.00  & 66.27$\pm$8.19  & 45.11$\pm$0.17   & 82.18$\pm$0.59  & 50.00$\pm$0.00  & 63.43$\pm$7.13  & 45.03$\pm$0.04   & 82.51$\pm$0.80  & 51.62$\pm$3.24  & 71.59$\pm$7.43  & 48.03$\pm$5.82  \\
ILRA-MIL          & 82.18$\pm$1.19  & 51.62$\pm$3.24  & 63.54$\pm$5.73  & 48.03$\pm$5.82   & 82.18$\pm$1.19  & 52.13$\pm$3.14  & 65.24$\pm$4.97  & 49.34$\pm$5.73   & 82.19$\pm$1.11  & 53.54$\pm$3.23  & 69.54$\pm$3.61  & 49.09$\pm$4.81   & 82.51$\pm$1.23  & 55.16$\pm$4.25  & 73.13$\pm$6.66  & 51.20$\pm$7.42  \\
ACMIL             & 79.87$\pm$3.63  & 56.16$\pm$3.28  & 73.30$\pm$7.24  & 55.85$\pm$5.61   & 79.89$\pm$4.56  & 58.92$\pm$4.98  & 74.61$\pm$6.06  & 60.06$\pm$6.43   & 80.20$\pm$4.52  & 72.56$\pm$7.67  & 71.84$\pm$3.38  & 60.34$\pm$6.66   & 82.21$\pm$2.85  & 62.62$\pm$3.50  & 77.40$\pm$4.97  & 64.70$\pm$7.43  \\
MambaMIL      & 82.52$\pm$2.14  & 53.75$\pm$5.65  & 77.35$\pm$5.11  & 52.41$\pm$9.63   & 79.89$\pm$3.41  & 58.89$\pm$3.45  & 76.74$\pm$5.28  & 60.06$\pm$5.55   & 81.86$\pm$2.65  & 56.88$\pm$4.18  & 75.67$\pm$5.90  & 57.34$\pm$7.37   & 84.83$\pm$1.82  & 66.86$\pm$4.37  & 84.26$\pm$1.68  & 68.87$\pm$2.46  \\
TDA-MIL           & 82.22$\pm$1.08  & 61.04$\pm$7.30  & 85.41$\pm$2.06  & 62.01$\pm$9.61   & 80.95$\pm$0.86  & 62.16$\pm$7.46  & 82.72$\pm$3.19  & 62.33$\pm$9.18   & 80.60$\pm$1.17  & 53.64$\pm$4.92  & 79.84$\pm$4.08  & 51.37$\pm$7.84   & 83.38$\pm$1.18  & 64.64$\pm$8.81  & 85.84$\pm$1.86  & 65.08$\pm$11.02 \\
\midrule
\textbf{Average}  & \underline{81.41} & \underline{53.36} & 69.77 & 50.73   & 80.97 & 53.39 & \underline{69.85} & 51.40   & 81.33 & 53.35 & 69.40 & \underline{50.50}   & \textbf{82.63} & \textbf{56.50} & \textbf{74.68} & \textbf{54.56} \\
\bottomrule
\end{tabular}%
}
\end{table}

\begin{table}[t!]
\centering
\caption{\footnotesize \textbf{Survival prediction performance across MIL architectures on TCGA-LUAD and TCGA-STAD.} The results are reported as mean$\pm$s.d. of held-out C-index under four preprocessing strategies. MUFASA consistently improved prognostic performance relative to CLAM-style preprocessing, Histolab, and Trident across pooling, attention-based, sequence-based, graph-based, and survival-specific MIL models. The best-performing preprocessor, averaged across models, is highlighted in bold, and the second-best is underlined (bottom row). }
\label{tab:5}
{\tiny
\resizebox{\textwidth}{!}{%
\begin{tabular}{l|cccc|cccc} 
\toprule
\multirow{2}{*}{\textbf{Model}}
    & \multicolumn{4}{c|}{\textbf{TCGA-LUAD}}
    & \multicolumn{4}{c}{\textbf{TCGA-STAD}} \\
\cmidrule(lr){2-5}\cmidrule(lr){6-9}
    & CLAM & Histolab & Trident & MUFASA
    & CLAM & Histolab & Trident & MUFASA \\
\midrule
Max-pooling       & 0.629$\pm$0.059 & 0.637$\pm$0.084 & 0.603$\pm$0.075 & 0.662$\pm$0.069   & 0.586$\pm$0.044 & 0.601$\pm$0.048 & 0.593$\pm$0.043 & 0.634$\pm$0.044 \\
Mean-pooling      & 0.628$\pm$0.051 & 0.627$\pm$0.055 & 0.626$\pm$0.058 & 0.642$\pm$0.049   & 0.623$\pm$0.058 & 0.623$\pm$0.061 & 0.619$\pm$0.064 & 0.639$\pm$0.068 \\
ABMIL             & 0.636$\pm$0.053 & 0.633$\pm$0.054 & 0.628$\pm$0.052 & 0.649$\pm$0.046   & 0.616$\pm$0.054 & 0.615$\pm$0.052 & 0.617$\pm$0.059 & 0.635$\pm$0.066 \\
CLAM-MB           & 0.630$\pm$0.064 & 0.629$\pm$0.068 & 0.622$\pm$0.068 & 0.655$\pm$0.061   & 0.612$\pm$0.072 & 0.612$\pm$0.073 & 0.608$\pm$0.069 & 0.641$\pm$0.085 \\
TransMIL          & 0.646$\pm$0.060 & 0.641$\pm$0.073 & 0.633$\pm$0.074 & 0.660$\pm$0.067   & 0.626$\pm$0.078 & 0.624$\pm$0.075 & 0.623$\pm$0.087 & 0.648$\pm$0.087 \\
DSMIL             & 0.634$\pm$0.062 & 0.631$\pm$0.061 & 0.632$\pm$0.066 & 0.663$\pm$0.059   & 0.611$\pm$0.071 & 0.612$\pm$0.073 & 0.605$\pm$0.073 & 0.643$\pm$0.079 \\
DTFD-MIL          & 0.627$\pm$0.053 & 0.627$\pm$0.054 & 0.628$\pm$0.061 & 0.653$\pm$0.045   & 0.621$\pm$0.054 & 0.620$\pm$0.056 & 0.626$\pm$0.060 & 0.659$\pm$0.067 \\
MambaMIL      & 0.624$\pm$0.047 & 0.623$\pm$0.047 & 0.616$\pm$0.056 & 0.639$\pm$0.041   & 0.616$\pm$0.045 & 0.613$\pm$0.047 & 0.623$\pm$0.053 & 0.637$\pm$0.057 \\
DeepAttnMISL  & 0.612$\pm$0.045 & 0.611$\pm$0.047 & 0.612$\pm$0.047 & 0.636$\pm$0.036   & 0.617$\pm$0.049 & 0.614$\pm$0.050 & 0.618$\pm$0.058 & 0.630$\pm$0.060 \\
OTSurv            & 0.647$\pm$0.065 & 0.645$\pm$0.064 & 0.647$\pm$0.069 & 0.664$\pm$0.058   & 0.632$\pm$0.061 & 0.632$\pm$0.062 & 0.632$\pm$0.060 & 0.661$\pm$0.067 \\
IBMIL             & 0.634$\pm$0.063 & 0.632$\pm$0.062 & 0.629$\pm$0.067 & 0.657$\pm$0.061   & 0.614$\pm$0.073 & 0.615$\pm$0.072 & 0.611$\pm$0.071 & 0.634$\pm$0.083 \\
ILRA-MIL          & 0.629$\pm$0.066 & 0.629$\pm$0.066 & 0.618$\pm$0.062 & 0.639$\pm$0.056   & 0.612$\pm$0.065 & 0.614$\pm$0.064 & 0.614$\pm$0.063 & 0.636$\pm$0.076 \\
PANTHER           & 0.638$\pm$0.070 & 0.633$\pm$0.065 & 0.633$\pm$0.065 & 0.655$\pm$0.062   & 0.612$\pm$0.071 & 0.619$\pm$0.073 & 0.608$\pm$0.073 & 0.638$\pm$0.085 \\
\midrule
\textbf{Average}  & \underline{0.6326} & 0.6306 & 0.6259 & \textbf{0.6518}   & 0.6152 & \underline{0.6165} & 0.6152 & \textbf{0.6411} \\
\bottomrule
\end{tabular}%
}}
\end{table}

\begin{table}[t!]
\centering
\caption{\footnotesize \textbf{Tumor diagnosis, tumor subtyping, and biomarker-status prediction using MUFASA-extracted Set1+Set2+Set3 tiles across MIL models.} Tumor diagnosis was evaluated on CAMELYON16, tumor subtyping on NSCLC, and biomarker-status prediction on the STANFORD and TCGA-CRC cohorts. Performance is reported as mean±s.d. across five repeated runs using different random seeds for CAMELYON16 and five-fold cross-validation for the other cohorts, based on accuracy (ACC), balanced accuracy (BACC), Macro-AUC, and Macro-F1.}
\label{tab:6}
\resizebox{1.4\textwidth}{!}{%
\begin{tabular}{l|cccc|cccc|cccc|cccc}
\toprule
\multirow{2}{*}{\textbf{Model}}
    & \multicolumn{4}{c|}{\textbf{CAMELYON16}}
    & \multicolumn{4}{c|}{\textbf{NSCLC}}
    & \multicolumn{4}{c|}{\textbf{STANFORD}}
    & \multicolumn{4}{c}{\textbf{TCGA-CRC}} \\
\cmidrule(lr){2-5}\cmidrule(lr){6-9}\cmidrule(lr){10-13}\cmidrule(lr){14-17}
    & ACC & BACC & Macro-AUC & Macro-F1
    & ACC & BACC & Macro-AUC & Macro-F1
    & ACC & BACC & Macro-AUC & Macro-F1
    & ACC & BACC & Macro-AUC & Macro-F1 \\
\midrule
Mean-pooling      & 62.48$\pm$2.00  & 53.86$\pm$4.35  & 61.00$\pm$3.16  & 46.94$\pm$9.76   & 82.07$\pm$1.02  & 81.68$\pm$1.16  & 89.54$\pm$1.22  & 81.99$\pm$0.98   & 86.40$\pm$0.28  & 60.39$\pm$2.02  & 82.51$\pm$2.24  & 62.91$\pm$2.52   & 81.46$\pm$1.53  & 50.42$\pm$1.79  & 68.42$\pm$3.63  & 46.50$\pm$3.13  \\
\underline{DeepAttnMISL}  & 62.01$\pm$0.00  & 50.00$\pm$0.00  & 69.49$\pm$1.01  & 38.27$\pm$0.00   & 70.60$\pm$4.19  & 69.21$\pm$3.93  & 77.52$\pm$3.03  & 69.35$\pm$4.43   & 84.88$\pm$0.22  & 50.00$\pm$0.00  & 74.04$\pm$1.53  & 45.91$\pm$0.06   & 82.18$\pm$0.59  & 50.00$\pm$0.00  & 61.06$\pm$7.36  & 45.11$\pm$0.17  \\
CLAM-MB           & 82.48$\pm$2.93  & 80.74$\pm$2.76  & 85.53$\pm$1.98  & 81.13$\pm$2.99   & 87.08$\pm$2.18  & 86.93$\pm$2.35  & 94.75$\pm$1.16  & 86.87$\pm$2.34   & 87.93$\pm$1.78  & 69.44$\pm$6.16  & 83.46$\pm$3.83  & 69.54$\pm$3.95   & 81.53$\pm$2.14  & 53.93$\pm$2.12  & 72.30$\pm$7.96  & 52.99$\pm$4.19  \\
DSMIL             & 70.18$\pm$6.91  & 61.44$\pm$9.60  & 71.83$\pm$3.32  & 57.83$\pm$16.16  & 77.11$\pm$1.84  & 75.61$\pm$1.91  & 83.09$\pm$2.96  & 76.06$\pm$1.81   & 83.69$\pm$1.06  & 53.28$\pm$2.67  & 74.71$\pm$4.28  & 52.36$\pm$4.5    & 82.18$\pm$0.59  & 50.00$\pm$0.00  & 68.48$\pm$6.77  & 45.11$\pm$0.17  \\
TransMIL          & 82.99$\pm$1.86  & 80.96$\pm$1.48  & 89.97$\pm$0.68  & 81.78$\pm$1.49   & 85.71$\pm$4.49  & 86.16$\pm$4.21  & 95.12$\pm$0.79  & 85.88$\pm$4.54   & 84.09$\pm$1.66  & 53.10$\pm$6.20  & 75.02$\pm$2.29  & 49.67$\pm$7.50   & 80.97$\pm$1.34  & 54.67$\pm$4.66  & 68.11$\pm$3.47  & 53.27$\pm$7.77  \\
DTFD-MIL          & 78.60$\pm$1.60  & 77.69$\pm$1.24  & 83.96$\pm$2.08  & 77.46$\pm$1.48   & 88.04$\pm$1.30  & 87.92$\pm$1.37  & 95.01$\pm$1.13  & 87.89$\pm$1.32   & 86.06$\pm$1.24  & 59.57$\pm$2.77  & 83.02$\pm$1.87  & 61.98$\pm$3.59   & 80.79$\pm$2.41  & 49.89$\pm$0.98  & 69.05$\pm$5.39  & 45.11$\pm$0.17  \\
MHIM-MIL          & 77.21$\pm$7.35  & 78.94$\pm$5.05  & 90.57$\pm$0.71  & 76.79$\pm$6.98   & 87.33$\pm$1.34  & 86.47$\pm$1.54  & 96.26$\pm$0.48  & 86.92$\pm$1.48   & 87.93$\pm$2.19  & 69.26$\pm$9.90  & 85.32$\pm$8.65  & 70.31$\pm$12.59  & 82.04$\pm$1.80  & 56.21$\pm$3.31  & 81.23$\pm$4.17  & 55.77$\pm$5.45  \\
Additive MIL      & 79.38$\pm$0.38  & 77.52$\pm$0.50  & 83.83$\pm$1.10  & 77.82$\pm$0.30   & 87.46$\pm$3.08  & 81.91$\pm$3.34  & 94.75$\pm$1.19  & 87.18$\pm$3.19   & 85.91$\pm$4.92  & 68.89$\pm$4.69  & 85.09$\pm$3.61  & 70.12$\pm$4.60   & 82.18$\pm$0.59  & 50.71$\pm$1.42  & 74.19$\pm$10.32 & 46.63$\pm$3.01  \\
DGR-MIL           & 68.38$\pm$9.33  & 58.52$\pm$12.58 & 76.48$\pm$13.65 & 51.22$\pm$18.0   & 58.36$\pm$11.92 & 56.50$\pm$13.00 & 85.77$\pm$6.35  & 43.71$\pm$18.54  & 70.24$\pm$27.52 & 59.17$\pm$7.71  & 83.41$\pm$1.88  & 52.36$\pm$21.42  & 79.87$\pm$2.46  & 50.82$\pm$1.00  & 68.26$\pm$7.13  & 44.94$\pm$0.29  \\
ILRA-MIL          & 62.01$\pm$0.00  & 50.00$\pm$0.00  & 68.53$\pm$2.14  & 38.27$\pm$0.00   & 88.35$\pm$2.60  & 88.01$\pm$2.03  & 95.16$\pm$0.70  & 88.28$\pm$2.20   & 87.33$\pm$0.89  & 70.94$\pm$4.59  & 86.07$\pm$3.04  & 72.25$\pm$3.55   & 82.18$\pm$0.59  & 50.00$\pm$0.00  & 68.59$\pm$2.81  & 45.11$\pm$0.17  \\
ACMIL             & 78.76$\pm$1.67  & 76.58$\pm$0.88  & 82.01$\pm$1.57  & 76.91$\pm$1.28   & 87.82$\pm$2.63  & 87.14$\pm$2.98  & 95.01$\pm$1.20  & 87.30$\pm$2.77   & 87.39$\pm$0.87  & 72.75$\pm$1.75  & 86.81$\pm$3.57  & 74.92$\pm$1.74   & 80.54$\pm$4.78  & 59.80$\pm$4.91  & 76.35$\pm$4.60  & 60.01$\pm$7.81  \\
\underline{MambaMIL}      & 78.61$\pm$8.89  & 76.58$\pm$9.86  & 80.40$\pm$9.57  & 76.83$\pm$9.82   & 87.65$\pm$2.32  & 87.26$\pm$2.62  & 94.98$\pm$0.52  & 87.34$\pm$2.49   & 88.66$\pm$1.63  & 75.51$\pm$4.81  & 89.27$\pm$3.79  & 76.51$\pm$3.71   & 82.85$\pm$2.99  & 60.42$\pm$7.48  & 83.74$\pm$2.78  & 60.20$\pm$7.41  \\
TDA-MIL           & 84.96$\pm$1.74  & 82.10$\pm$2.34  & 90.03$\pm$0.86  & 83.22$\pm$2.09   & 87.81$\pm$4.98  & 87.40$\pm$6.01  & 96.36$\pm$0.89  & 87.27$\pm$5.87   & 88.26$\pm$1.00  & 73.71$\pm$3.29  & 90.06$\pm$2.84  & 75.30$\pm$2.05   & 83.18$\pm$1.54  & 59.72$\pm$6.31  & 84.10$\pm$3.25  & 60.10$\pm$8.51  \\
\midrule
\textbf{Average}  & 74.46 & 69.61 & 79.57 & 66.49   & 82.70 & 82.10 & 91.79 & 81.21   & 85.29 & 64.30 & 82.98 & 64.16   & \textbf{81.68} & \textbf{53.58} & \textbf{72.60} & \textbf{50.83} \\
\bottomrule
\end{tabular}%
}
\end{table}

\end{landscape}

\begin{table}[t!]
\centering
\caption{\footnotesize \textbf{Survival prediction performance across MIL models using Set1+Set2+Set3 tiles extracted by MUFASA.} Performance is reported as mean$\pm$s.d. of the held-out C-index.}
\label{tab:7}
{\scriptsize
\begin{tabular}{l|cc}
\toprule
\textbf{Model} & \textbf{MUFASA (LUAD)} & \textbf{MUFASA (STAD)} \\
\midrule
Max-pooling           & 0.635$\pm$0.075 & 0.587$\pm$0.044 \\
Mean-pooling          & 0.638$\pm$0.050 & 0.615$\pm$0.069 \\
ABMIL                 & 0.644$\pm$0.055 & 0.609$\pm$0.076 \\
CLAM-MB               & 0.633$\pm$0.061 & 0.605$\pm$0.084 \\
TransMIL              & 0.645$\pm$0.067 & 0.621$\pm$0.087 \\
DSMIL                 & 0.633$\pm$0.056 & 0.605$\pm$0.079 \\
DTFD-MIL              & 0.637$\pm$0.046 & 0.616$\pm$0.065 \\
MambaMIL  & 0.628$\pm$0.040 & 0.617$\pm$0.057 \\
DeepAttnMISL & 0.626$\pm$0.036 & 0.611$\pm$0.058 \\
OTSurv                & 0.654$\pm$0.060 & 0.634$\pm$0.071 \\
IBMIL                 & 0.639$\pm$0.059 & 0.611$\pm$0.081 \\
ILRA-MIL              & 0.633$\pm$0.055 & 0.612$\pm$0.074 \\
PANTHER               & 0.641$\pm$0.061 & 0.610$\pm$0.082 \\
\midrule
\textbf{Average}      & \textbf{0.638}  & \textbf{0.611}  \\
\bottomrule
\end{tabular}
}
\end{table} 

\renewcommand{\thefigure}{S\arabic{figure}}
\setcounter{figure}{0}  

\begin{figure}[t!] 
	\centering 
	\includegraphics[width=0.95\textwidth]{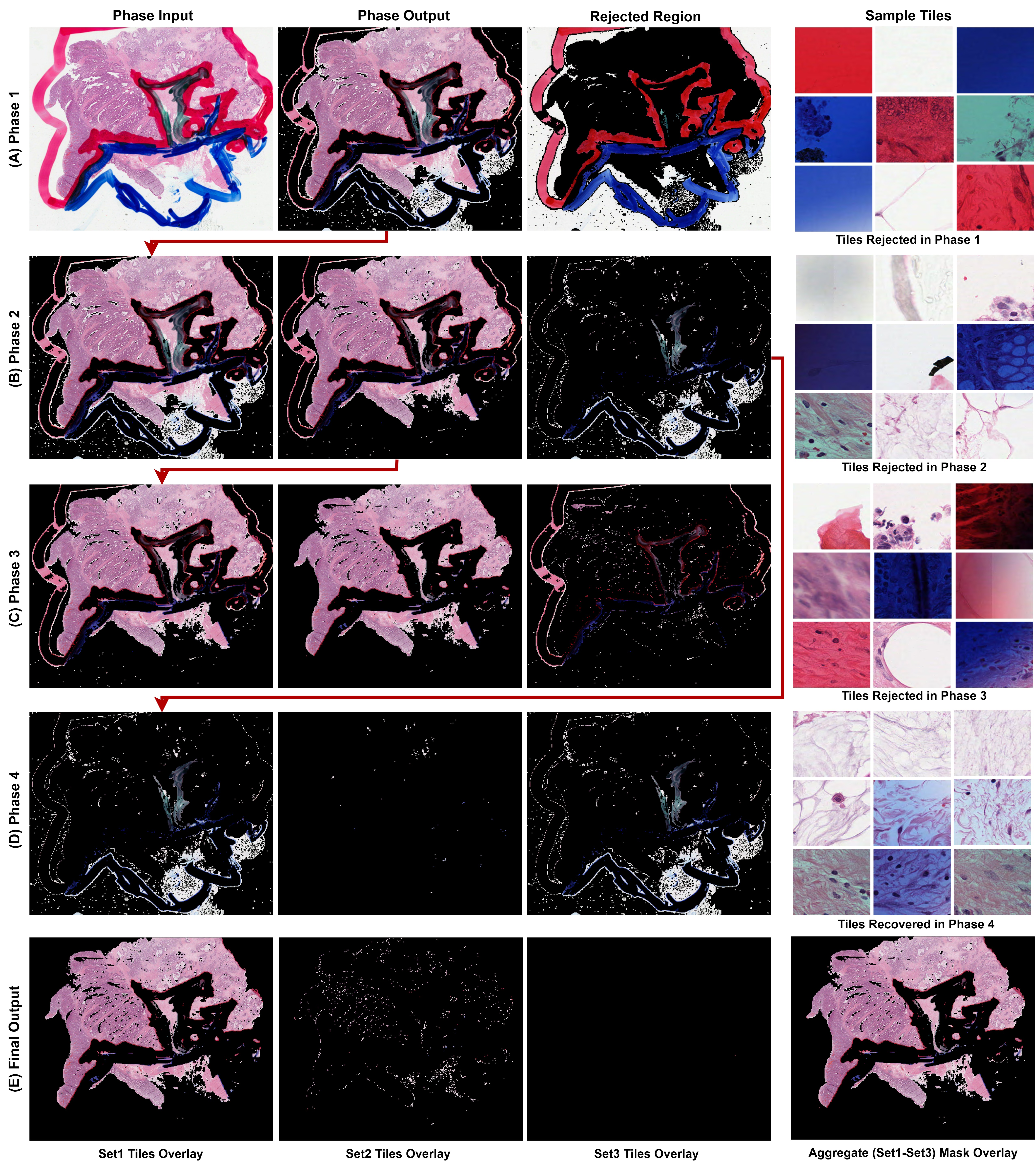} 
	\caption{\footnotesize \textrm{\textbf{Phase-wise audit trail of MUFASA preprocessing: each phase input and output, rejected region, and representative tiles rejected. A. }Input WSI from TCGA-COAD (slide ID: TCGA-F4-6809-01Z-00-DX1.5ab8333f-0c77-4685-8701-4130a93e6f3a) showing representative artifact patterns commonly encountered in routine histopathology, including background-dominant regions, pen marks, blur, dust, and staining irregularities. Phase 1 removes slide-level artifacts, primarily pen marks and associated high-contrast ink regions, before tile extraction. 
    \textbf{B.} Phase 2 applies optical-density and basic quality filtering to suppress background-dominant, faintly stained, and low-contrast artifact tiles. 
    \textbf{C.} Phase 3 performs reconstruction-based utility filtering to remove residual subtle artifacts, including blur, dust or debris, and low tissue content regions that evade deterministic thresholds. 
    \textbf{D.} Phase 4 selectively recovers biologically relevant but visually subtle tissue types, such as adipose or mucin, from the Phase 2 rejected pool. 
    \textbf{E.} Spatial overlays of the final MUFASA output, showing tile stratification into Set1, Set2, and Set3, remapped to their original WSI coordinates, along with the combined full mask overlay. }}
	\label{fig:S1}       
\end{figure} 

\begin{figure}[t!] 
	\centering 
	\includegraphics[width=0.95\textwidth]{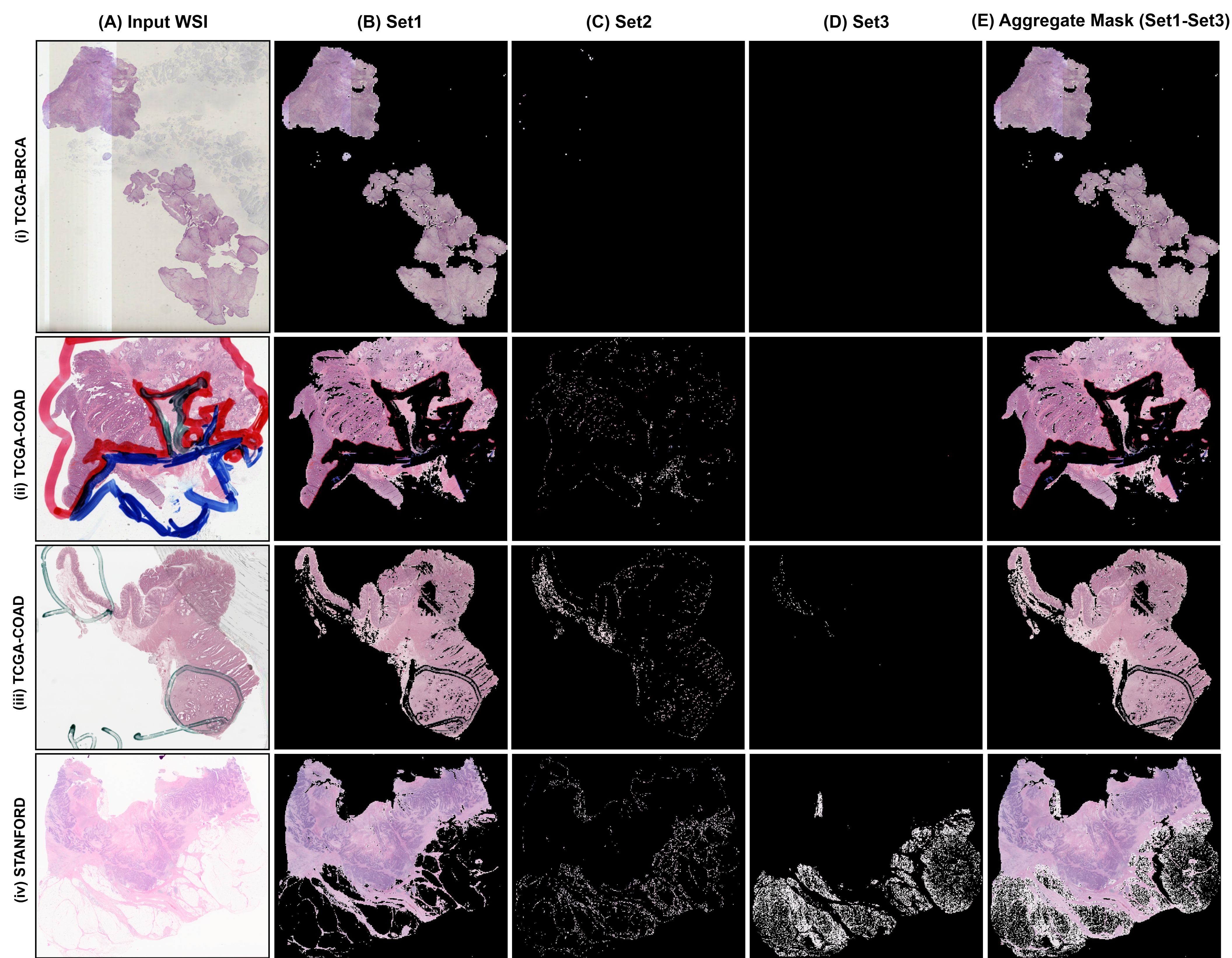} 
	\caption{\footnotesize \textrm{\textbf{Utility-aware tissue mask outputs generated by MUFASA across colorectal and breast cancer cohorts. A.} Input WSI. Examples are shown from TCGA-BRCA (i, ii), TCGA-COAD (iii), and STANFORD (iv) CRC. 
    \textbf{B–D.} Category-specific tissue masks obtained by spatially remapping tiles assigned to Set1, Set2, and Set3, respectively, to their original slide coordinates. 
    \textbf{E.} Aggregate tissue mask (Set1-Set3). The overlays show that MUFASA suppresses artifact-dominant and low-utility regions while preserving tissue-containing areas across heterogeneous slides and cohorts. Slide IDs: (i) TCGA-AC-A5EI-01Z-00-DX1.A174D91A-730E-460C-AE73-46CA1E5B177B, (ii) TCGA-F4-6809-01Z-00-DX1.5ab8333f-0c77-4685-8701-4130a93e6f3a, and (iii) TCGA-DM-A1D4-01Z-00-DX1.38346604-2BAF-44F8-BD96-5BF58253C6AD.
 }}
	\label{fig:S2}       
\end{figure}

\end{document}